\documentclass{aastex63}

\usepackage{graphicx} 
\usepackage[sort&compress]{natbib}   
\usepackage{amsmath,amssymb}
\shorttitle{Spec-S5: The Spectroscopic Stage-5 Experiment}
\shortauthors{Spec-S5 Collaboration}

\newcommand{\hmpcC}{h^3~{\rm Mpc}^{-3}}
\defcitealias{p5report2014}{P5 2014}
\defcitealias{p5report2023}{P5 2023}

\begin{document}

\title{The Spectroscopic Stage-5 Experiment}

\correspondingauthor{Claire Poppett }
\email{clpoppett@berkeley.edu}
\author{Robert~Besuner}
\affiliation{Lawrence Berkeley National Laboratory, 1 Cyclotron Rd., Berkeley CA 94720}
\author{Arjun~Dey}
\affiliation{NSF NOIRLab, 950 N. Cherry Ave., Tucson, AZ 85719}
\author{Alex~Drlica-Wagner}
\affiliation{Fermi National Accelerator Laboratory, P.O.\ Box 500, Batavia, IL 60510, USA}
\affiliation{Kavli Institute for Cosmological Physics, University of Chicago, Chicago, IL 60637, USA}
\affiliation{Department of Astronomy and Astrophysics, University of Chicago, Chicago, IL 60637, USA}
\author{Haruki~Ebina}
\affiliation{Department of Physics, University of California, Berkeley, CA 94720}
\author{Guillermo~Fernandez~Moroni}
\affiliation{Fermi National Accelerator Laboratory, P.O.\ Box 500, Batavia, IL 60510, USA}
\author{Simone~Ferraro}
\affiliation{Lawrence Berkeley National Laboratory, 1 Cyclotron Rd., Berkeley CA 94720}
\author{Jaime~E.~Forero-Romero}
\affiliation{Observatorio Astron\'omico, Universidad de los Andes, Cra. 1 No. 18A-10, CP 111711 Bogot\'a, Colombia}
\author{Klaus~Honscheid}
\affiliation{Department of Physics, Ohio State University, Columbus, OH 43210, USA}
\author{Patrick~Jelinsky}
\affiliation{Space Sciences Laboratory, University of California, Berkeley, 7 Gauss Way, Berkeley, CA  94720, USA}
\affiliation{Lawrence Berkeley National Laboratory, 1 Cyclotron Rd., Berkeley CA 94720}
\author{Dustin~A.~Lang}
\affiliation{Perimeter Institute for Theoretical Physics, 31 Caroline St.~N, Waterloo, ON N2L-2Y5, Canada}
% ...

\begin{abstract}
The existence, properties, and dynamics of the dark sectors of our universe pose fundamental challenges to our current model of physics, and large-scale astronomical surveys may be our only hope to unravel these long-standing mysteries. In this white paper, we describe the science motivation, instrumentation, and survey plan for the next-generation spectroscopic observatory, the Stage-5 Spectroscopic Experiment (Spec-S5). Spec-S5 is a new all-sky spectroscopic instrument optimized to efficiently carry out cosmological surveys of unprecedented scale and precision. The baseline plan for Spec-S5 involves upgrading two existing 4-m telescopes to new 6-m wide-field facilities, each with a highly multiplexed spectroscopic instrument capable of simultaneously measuring the spectra of 13,000 astronomical targets. Spec-S5, which builds and improves on the hardware used for previous cosmology experiments, represents a cost-effective and rapid approach to realizing a more than 10$\times$ gain in spectroscopic capability compared to the current state-of-the-art represented by the  Dark Energy Spectroscopic Instrument project (DESI). Spec-S5 will provide a critical scientific capability in the post-Rubin and post-DESI era for advancing cosmology, fundamental physics, and astrophysics in the 2030s. 
\end{abstract}

\section{Introduction}

Observational cosmology has provided an extraordinary perspective on our universe and our place within it.
However, as our understanding of the universe has increased, some glaring holes in our knowledge have become apparent: What physics is responsible for the super-luminal expansion of the universe at early times? What drives the accelerating expansion of the universe at late times? What is the nature of the mysterious dark matter that makes up 83\% of the matter in the universe? These fundamental physical questions are intimately linked to fundamental astronomical questions about how galaxies and stars have formed and evolved within our universe. Cosmic surveys are the primary means by which we study the origin, structure, composition, and evolution of our universe.
In particular, many of these questions require the spectroscopy of large numbers of astronomical sources. Spectroscopy provides key data on the physics of primordial inflation and late-time cosmic acceleration, the astrophysics of galaxy evolution, and the nature and effects of dark matter in galaxies. For example, observable features in the three-dimensional structure of the universe are key predictions of cosmological models, and their time evolution provides unique constraints on the nature of dark energy, inflation and the gravitational effects of dark matter. The next major advance in our understanding of the universe requires spectroscopic measurements of hundreds of millions of astronomical objects. 


Here, we present the plans for the next-generation spectroscopic observatory, the Stage-5 Spectroscopic Experiment (Spec-S5),\footnote{\url{https://spec-s5.org}} where this name recognizes its heritage from previous generations of cosmic surveys \citep[cf.][]{DETF_Report_2006}.  Spec-S5 will be an exceptionally powerful cosmology experiment within the High Energy Physics Cosmic Frontier research program, mapping the universe in three dimensions and studying its evolution with time. Designed to access the entire sky and comprising two instruments each of which will efficiently measure more than ten thousand spectra at a time, Spec-S5 represents more than a factor of 10 increase in both survey speed and information content relative to the current state-of-the-art. Spec-S5 will address frontier problems in fundamental physics and the growth of structure in the universe. The all-sky coverage of Spec-S5 will allow it to uniquely study the structure and formation history of the entire Milky Way as well as provide the best constraints on the lowest-order modes of the galaxy power spectrum. The data from Spec-S5 will complement and enhance other current and future observatories, including the Vera C.\ Rubin Observatory's Legacy Survey of Space and Time \citep[LSST;][]{2019ApJ...873..111I} and the Stage 4 Cosmic Microwave Background experiment \citep[CMB-S4;][]{abazajian2016cmbs4sciencebookedition}, to boost the reach of each experiment.  

Spec-S5 builds upon the heritage of generations of cosmology experiments that have conducted redshift surveys.  The Baryon Oscillation Spectroscopic Survey (BOSS), which operated from 2009--2014, upgraded the instrumentation on the 2.5-m Sloan Telescope to map 1.25 million galaxies and 386,000 quasars \citep{BOSS_2012,BOSS_2013}.  A continuation of this program, the extended BOSS (eBOSS), operated from 2014--2019 and mapped an additional 530,000 galaxies and 530,000 quasars \citep{eBOSS_2016}.  In combination, BOSS and eBOSS resulted in the first Stage 3 dark energy constraints using the baryon acoustic oscillation (BAO) feature \citep{eBOSS_2021}.  The Dark Energy Spectroscopic Instrument \citep[DESI;][]{desi_inst,DESI2016_Part1} on the 4-m Mayall Telescope is currently (i.e., 2021-26) assembling a map of 40 million galaxies and quasars.  DESI released the first dark energy constraints from a Stage-4 experiment in 2024, tantalizingly suggesting an evolving equation of state and providing the strongest constraints to date on modified gravity models \citep[e.g.,][]{DESI_FullShape_2024,DESI_BAO_Cosmology_2024}.  This sequence of experiments has demonstrated the power of increasingly capable redshift surveys as cosmological probes; Spec-S5 is the necessary next step.

The Spec-S5 reference design upgrades two existing National Science Foundation (NSF) facilities to provide access to the entire sky at the depth and multiplex required to achieve the full science reach of a Stage-5 experiment.  These two facilities currently host prior-generation experiments built and funded by the Department of Energy (DOE) Office of Science (OS): DESI \citep{desi_inst} at the Nicholas U.\ Mayall Telescope at Kitt Peak National Observatory in Arizona, and the Dark Energy Camera \citep[DECam;][]{DECam2015} at the Victor M.\ Blanco Telescope at Cerro Tololo Inter-American Observatory in Chile. Spec-S5 builds upon the experience of the enormously successful DESI and DECam projects. 
The reference design requires optical upgrades of the two telescope platforms, replacing the 4-m primary mirrors with new 6-m primary mirrors that each deliver a 2.2~deg diameter field of view. The focal plane will be populated with miniaturized robotic fiber positioners that improve on the DESI designs and provide more robust operations. The spectrographs are duplicates of those used by DESI with upgraded, lower-noise detectors. Much of the remaining hardware will be similar to that demonstrated by DESI, which makes this approach low-risk and cost-effective. 
%
%

Between now and the start of Spec-S5, DESI will complete its initial map of 40 million galaxies and quasars and deliver bright-time spectroscopic measurements of more than 12 million stars.  A DESI-2 program has been recommended by the 2023 Particle Physics Project Prioritization Panel (P5) report\footnote{The P5 Report was commissioned by the High Energy Physics Advisory Panel (HEPAP) which advises the US Federal Government on the national program in experimental and theoretical high energy physics.} to map at least an additional 10 million galaxies, expanding DESI's impact on dark energy science while also demonstrating the experimental approach for Spec-S5 with surveys of the early ($z>2$) universe to study primordial physics and place astrophysical constraints on the fundamental nature of dark matter \citep{p5report2023}. Spec-S5 will build upon these surveys and strong collaborations to deliver the most cost-effective and rapid approach to a Stage-5 survey \citep{schlegel2022spectroscopicroadmapcosmic}. These spectroscopic observations will be an exceptional complement to contemporaneous imaging surveys from the European Euclid satellite, Rubin LSST, and NASA's {\it Nancy Grace Roman Space Telescope}.
DESI and DESI-2 will operate concurrently with a number of other spectroscopic redshift surveys: Sumire/PFS on the Subaru Telescope, 4MOST at ESO/Vista, WEAVE at WHT, and Euclid.  The redshift surveys from these experiments are smaller or comparable in scale to Stage-4 experiments, and therefore will not represent significant additional scientific reach beyond DESI. 
Finally, we do not expect new experimental techniques such as line intensity mapping (LIM) to have reached a similar level of sensitivity by the end of the coming decade \citep[e.g.,][]{p5report2023}.

Spec-S5 provides a critical scientific capability for the coming decades and an order-of-magnitude leap beyond the current state-of-the-art. 
Several recent high-level astronomy and physics community reviews have consistently identified the necessity for enhanced multi-object spectroscopic capabilities. The Astro2020 Decadal Survey emphasized ``highly multiplexed spectroscopy" as a key priority, particularly through the NSF Astronomy Mid-Scale Program, which it designated as the ``highest priority sustaining activity from the ground" \citep{Astro2020}. The 2023 P5 Report \citep{p5report2023} specifically endorsed Spec-S5, recommending R\&D funding across all budget scenarios and construction under favorable conditions. Subsequently, the HEPAP Subcommittee on Facilities evaluated the refined Spec-S5 design and concluded that it was both ``absolutely central" to the field and ``ready to initiate construction". The advent of Rubin/LSST will greatly increase the need for supporting spectroscopy, making Spec-S5 a critical facility for the US astronomical community in the Rubin and post-LSST era. 

In this white paper, we briefly review the planned Spec-S5 experiment. In Section \S~\ref{sec:Science} we discuss the main scientific drivers, briefly reviewing the contributions to our understanding of dark energy, inflation, and dark matter made possible by a Stage 5 experiment. 
In Section~\S~\ref{sec:Landscape} we briefly review the landscape of current and future spectroscopic redshift surveys.
In Section~\S~\ref{sec:Survey} we outline a ``reference survey'' that would accomplish the science, describing the footprint, target selection, and survey plan.
Section \S~\ref{sec:Timeline} presents a project timeline, and Section \S~\ref{sec:Synergies} synergies with other facilities.
Section \S~\ref{sec:Instrument} describes the end-to-end instrument design. We summarize in Section \S~\ref{sec:Summary}.

\section{Science Motivation}
\label{sec:Science}

Spec-S5 will enable unprecedented, all-sky spectroscopic coverage that will revolutionize many different branches of astronomy, cosmology, and physics. Here, we describe fundamental questions in three broad areas that motivate the key science projects and drive top-level design, including aperture size, multiplexing, and spectral resolution. However, the data collected in pursuit of these key project measurements will enable a myriad of science projects well beyond the ones we describe here \citep[cf.\ the low-resolution science cases for the Maunakea Spectroscopic Explorer (MSE),  Wide-field Spectroscopic Telescope (WST), and Multiplexed Survey Telescope (MUST);][]{MSE_Science,WST_Science2024,MUST2024}, as demonstrated by the broad impact of the SDSS and DESI projects. 

\subsection{Science Driver 1: Fundamental Physics at $z>2$}

The matter distribution in the high-redshift universe ($z>2$) provides some of the most readily accessible information about the extreme conditions in the early universe, conditions that enable some unique probes of fundamental physics. The most fundamental of these is the nature of primordial inflation, although these data also contain valuable information about other key topics in physics, such as the number of relativistic species, the sum of neutrino masses, and evidence for early dark energy. The potential of massively multiplexed redshift surveys to address these questions are thoroughly discussed in the Snowmass white paper on primordial physics \citep{ferraro22} and further explored in \cite{Sailer_2021}, \cite{Cabass_2023} and \cite{Ebina_2024}, with the latter methodology adopted for this forecasting.
The measurements that provide information about primordial inflation include measurements of the local form of primordial non-Gaussianity ($f^\mathrm{local}_\mathrm{nl}$), which will help determine if inflation was driven by one or multiple fields, and departures from scale invariance that would provide more information about the inflationary potential or other possibilities such as axion-like particles or other particle production. Better measurements of the power spectrum will improve the precision on the sum of the neutrino masses and effective number of relativistic species, $N_\mathrm{eff}$, that will provide tests for new physics beyond the Standard Model of particle physics. Furthermore, baryon acoustic oscillation (BAO) and redshift space distortion (RSD) measurements at these early times will enable precision tests of early dark energy models and of the growth of structure in the matter-dominated regime, where the latter is relevant for understanding the ongoing $\sigma_8$ tension \citep[e.g.][]{Abdalla_2022, Beltr_n_Jim_nez_2025}.

DESI is currently using three-dimensional clustering of quasars and absorption in the Lyman-$\alpha$ forest of quasars to measure BAO at high redshifts ($z>2$), while the recently launched Euclid satellite \citep{laureijs2011eucliddefinitionstudyreport} will complement these measurements to redshifts $z\sim2$ \citep{euclidoverview}.  However, at redshifts $z>2.1$, 
these ongoing DESI and Euclid surveys only sample a relatively small fraction of the effective volume available to studies of large-scale structure.  The Spec-S5 reference survey (\S~\ref{sec:Survey}) is primarily motivated by the need to map an enormous volume of the universe, which is only possible by extending to higher redshift and by all-sky coverage. 

The information content available in large-scale structure can be quantified through the primordial figure of merit (PFoM), as defined in the Snowmass white paper on primordial physics \citep{ferraro22}.  Over all redshifts, DESI is forecast to obtain PFoM $\sim$ 0.9.  Spec-S5 is designed to increase the information content by an order of magnitude relative to DESI, indicated by a PFoM $\gtrsim$ 9.  This program can be achieved with a survey covering 11,000 deg$^2$ using 62 million galaxy targets selected to span the redshift range $2.1<z<4.5$.  Making a relatively conservative assumption of galaxy bias,
we can achieve a ${\rm PFoM} = 9.86$ using galaxies at a comoving number density of $8.0 \times 10^{-4}\,\hmpcC$ over the redshift range $2.1<z<3.5$ and $2.0\times 10^{-4}\,\hmpcC$ over the redshift range $3.5<z<4.5$.

\begin{figure}[t]
    \centering
    \includegraphics[width=6in]{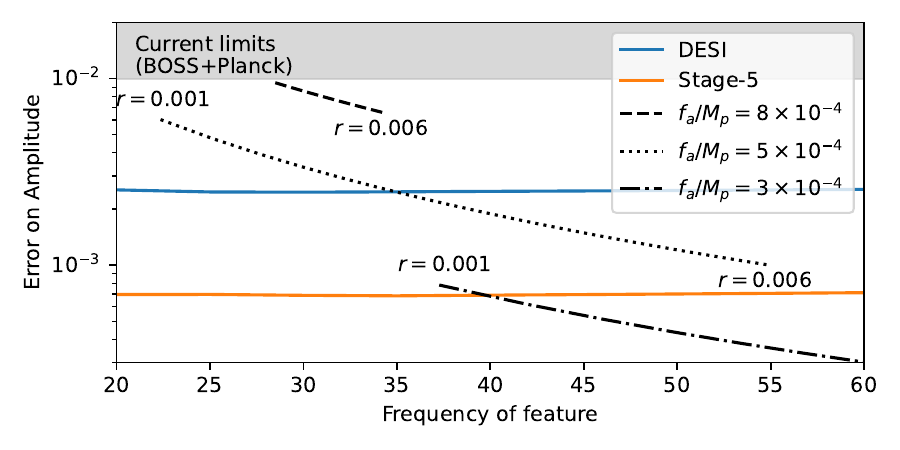}
    \caption{
    Predicted precision of Spec-S5 measurements on the amplitude of primordial features ($A_{\rm log}$) as a function of frequency ($\omega_{\rm log}$).
    The three models presented as black curves represent inflationary models that are inspired by string theory and predict both primordial features and primordial gravitational waves (denoted by $r$, the tensor-to-scalar ratio).  
    Specifically, an axion monodromy inflationary model \citep[e.g.][]{mcallister10,flauger10} is assumed with axion scale as listed in the legend.
    }
    \label{fig:features}
\end{figure}

\begin{figure}[thb]
    \centering
    \includegraphics[width=6.5in]{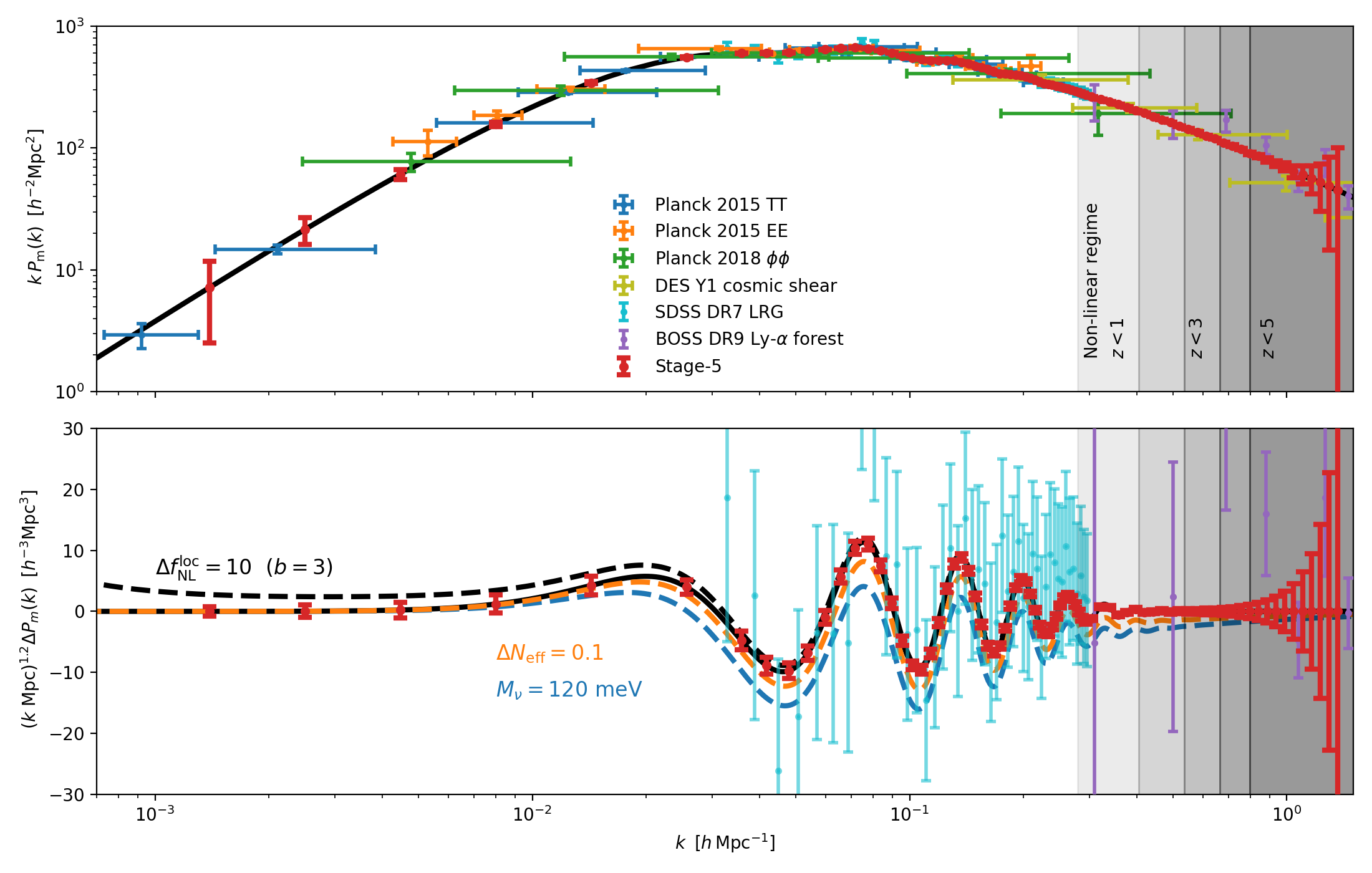}
    \caption{
    {\bf Top:}  Measurements of the power spectrum resulting from a wide array of CMB and galaxy redshift surveys. Spec-S5 (red points) represents a significant improvement on existing (and planned) measurements, resulting in very precise constraints across a wide range of $k$. 
    {\bf Bottom:}  The normalized power spectrum, showing the small deviations that result from the presence of non-Gaussianity (black dashed line), a massive neutrino (blue dashed line), or light relic particles (yellow dashed line).  Spec-S5 has the potential to measure these small deviations out to $k\approx0.7$ in the matter-dominated, high-redshift universe.
    }
    \label{fig:powerspectrum}
\end{figure}

New, high-precision measurements of large-scale structure spanning distance, growth, and inflationary physics will be possible with a Spec-S5 program designed to this specification.  The Spec-S5 reference survey described in \S~\ref{sec:Survey} will allow a holistic study of fundamental physics, as follows (see also Figure \ref{fig:powerspectrum}):
\begin{itemize}

\item {\bf BAO measurements at better than 0.1\% precision over the redshift range $\boldsymbol{2.1 < z < 4.5}$.}  A new measurement of the BAO parameter at these redshifts will allow tests of dark energy models that mimic a cosmological constant at low redshift or otherwise cause deviations in the expansion history at early times and thus evade detection by the lower redshift, Stage 4 dark energy probes.  Given current capabilities, only BAO measurements from a large spectroscopic sample can be used to constrain the expansion history in this era.  The impact of this BAO measurement relative to the dark energy content is described in the supporting Snowmass paper \citep{ferraro22} and presented in Figure~\ref{fig:hubble_diagram}.

\begin{figure}[thb]
    \centering
    \includegraphics[width=6in]{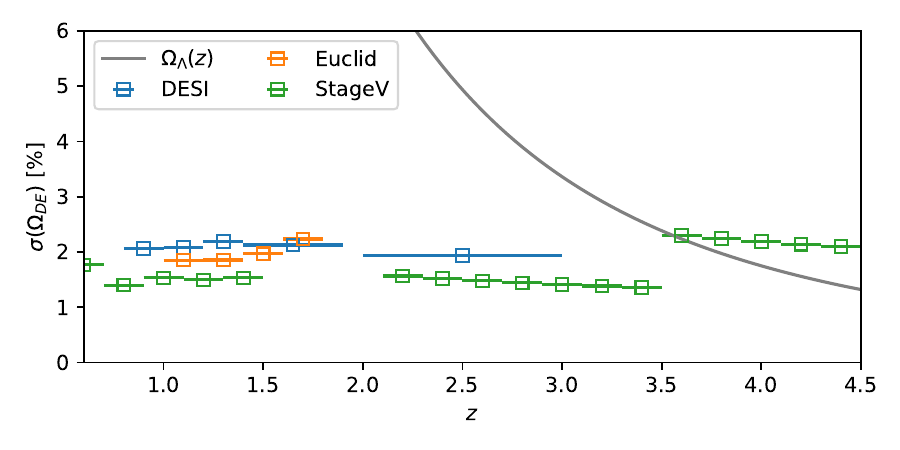}
    \caption{
    Predicted precision on the absolute error on the dark energy fraction ($\Omega_{\rm DE}$) as a function of redshift.
    The solid curve presents the relative energy density of dark energy under the fiducial $\Lambda$CDM model.
    }
    \label{fig:hubble_diagram}
\end{figure}

\item {\bf RSD measurements on the growth of structure at 0.55\% precision over the redshift range $\boldsymbol{2.1 < z < 4.5}$.}
Sub-percent measurements of the growth of structure in this redshift range, deep into matter-domination, will shed important light on the current ``$S_8$ tension'' in a new regime while allowing us to disentangle the effect of dark energy (primarily acting at $z < 1$), from that of massive neutrinos or modifications to general relativity on large scales. These constraints represent an order of magnitude improvement over current ones.

\item {\bf Measurements of $\boldsymbol{N_{\rm eff}}$ to a precision of 0.040 independent of CMB constraints.}  By constraining the thermal history of the universe over all times back to an epoch very close to the end of inflation, these measurements will provide tests for new relativistic species that are not currently included in the standard model.

\item {\bf Constraints on the summed mass of the neutrino mass eigenstates to a precision of 28 meV independent of low redshift measurements.}  The strongest cosmological constraints on neutrino mass will come from the combination of CMB-S4, low-redshift BAO and RSD measurements, and Stage-5 measurements.  We expect final constraints on neutrino mass with a precision that would allow us to exclude the inverted hierarchy at several $\sigma$ confidence in a minimal mass scenario.   

\item {\bf Constraints on the local form of primordial non-Gaussianity to a precision $\boldsymbol{\sigma(f_{\rm nl}^{\rm local})=1.4}$.}  Such measurements are necessary to determine if the period of inflation was driven by a single field or multiple fields, as well as to allow studies of excited states within the inflationary field ($f_{\rm nl}^{\rm orthogonal}$) and searches for interactions between the inflation field and particles present at the time of inflation ($f_{\rm nl}^{\rm equilateral}$).

\item {\bf Search for departures from scale invariance in the inflationary field at a precision that exceeds DESI forecasts by a factor of 4.}  Several theoretically well-motivated physical mechanisms can create features in the power spectrum—e.g., a phase transition in the inflationary potential, $<10^{-22}$ eV mass axion-like particles, early universe particle production, or other periodic corrections to the inflationary potential \citep{Slosar2019}. As one example, the sensitivity of Spec-S5 to some oscillatory primordial features is demonstrated in Figure~\ref{fig:features}.

\end{itemize}

\subsection{Science Driver 2: Dark Energy}

The Stage 4 dark energy experiments were motivated by the late-time transition to an accelerating universe driven by dark energy, as recommended by the 2014 P5 report \citep{p5report2014}. While the CMB measurements consistently showed excellent agreement with a $\Lambda$CDM model, the 2024 results from the first year of DESI data indicated a preference for time evolution of the dark energy equation of state \citep{DESI_BAO_Cosmology_2024}, at the $2.6\sigma$ level for DESI BAO and CMB, and at a level ranging from $2.5\sigma$ to $3.9\sigma$ when further including supernovae data from Pantheon+ \citep{brout22}, Union 3 \citep{rubinetal2023} or DES-SN5YR \citep{DESY5SN}.  Parameterizing the dark energy equation of state as $w(a) = w_0+w_a (1-a)$ (where $a=1/(1+z)$ is the scale factor associated with cosmic expansion), 
the combination of DESI and other cosmological data prefers models where $w_0>-1$ and $w_a<0$, implying that the dark energy density is evolving with time.
With these hints for dynamic dark energy, it becomes increasingly important to ensure a large and systematic study of dark energy at all redshifts. The sensitivity of DESI is excellent at $z\lesssim 2$, but limited by the density of bright quasars for the measurement in the Ly$\alpha$ at higher redshifts.  The Spec-S5 $z>2$ survey proposed here would remedy this issue by using large numbers of galaxy tracers in addition to probing the Ly$\alpha$ forest with fainter (and more numerous) QSOs.

\begin{figure}[t]
    \centering
    \includegraphics[width=6.5in]{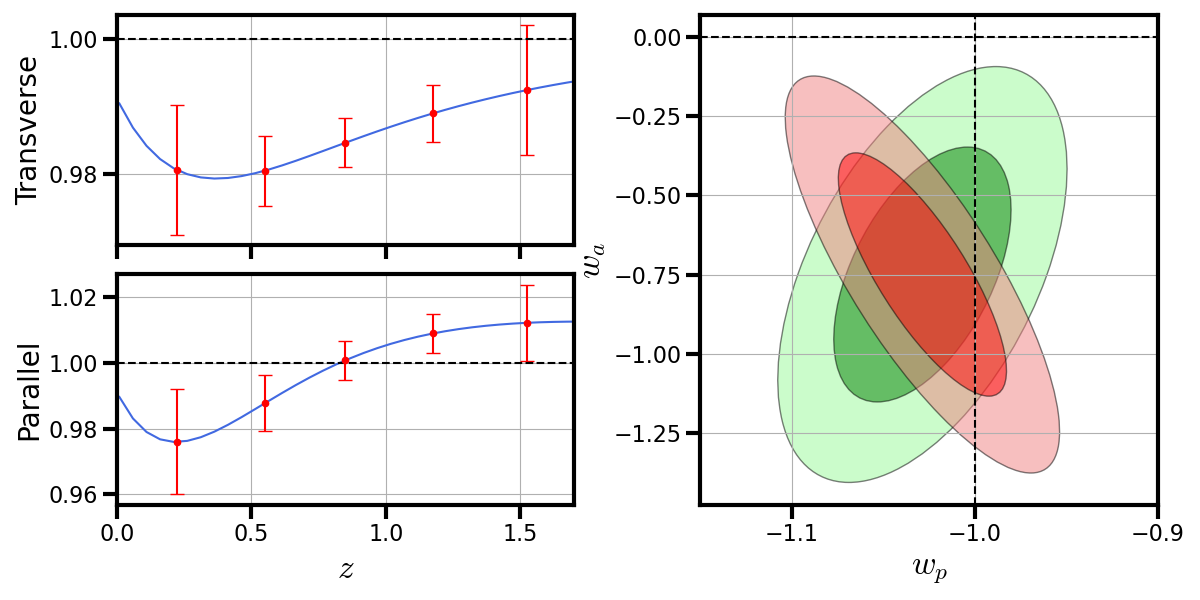}
    \caption{
    {\bf Left:}  The deviation in the position of the BAO feature relative to a $\Lambda$CDM model in the parallel and transverse directions as predicted by the preferred model of \cite{DESI_BAO_Cosmology_2024}, i.e.\ $w_0=-0.927$ and $w_a=-0.75$.  Overlaid on the preferred model (solid line) are the forecasted BAO measurements and error bars as would be expected from a 25,000 deg$^2$ program.
    {\bf Right:}  Precision on the parameters for the time-varying dark energy equation of state presented in a plane of $w_0-w_p$ with a pivot redshift $z_p=0.37$ chosen as a compromise between the BAO and SNe~Ia samples.  
    The green contours represent the 68\% and 95\% confidence intervals from the 2024 DESI results \citep[DESI BAO, Planck CMB, Planck+ACT CMB lensing, Pantheon+;][]{DESI_BAO_Cosmology_2024}.
    The red contours represent the 68\% and 95\% confidence intervals when combining Planck CMB and Spec-S5 BAO measurements over a 25,000 deg$^2$ footprint, excluding the complementary supernova data sets.
    }
    \label{fig:25k}
\end{figure}

DESI is currently being used to explore dark energy through measurements of BAO and RSD in the clustering of galaxies out to redshifts $z<1.$6, QSOs as tracers to $z<2.1$, and BAO with the Ly$\alpha$ forest at higher redshift. While the current data tantalizingly hint at a dynamical dark energy, only the full-sky coverage and large target density of Spec-S5 provide the necessary precision to fully explore the time evolution of the equation of state. 
Spec-S5 would obtain measurements of the isotropic BAO feature with a cumulative precision of 0.18\% at $z<1.1$, 0.23\%  over the interval $1.1<z<1.9$,  0.19\%  over the interval $2.1<z<3.5$, and 0.22\% over the interval $3.5<z<4.5$. It will therefore not only lead to an improvement by close to a factor of 2 over the forecasts for the existing DESI program at high redshifts  \citep{SV}, but also enable the study of dark energy with a similar constraining power up to  $z= 4.5$. Even more precise results will be obtained when adding constraints from the cross-correlation of this data set with other tracers such as the Ly$\alpha$ forest. In turn, these measurements, when combined with Planck CMB data, will allow supernova-independent tests that can discriminate between $\Lambda$CDM and the preferred central value of the Pantheon+ SN program \citep{brout22} at $>5\sigma$ confidence, as shown in Figure~\ref{fig:25k}.

\subsection{Science Driver 3: Dark Matter}

The fundamental nature of dark matter remains one of the outstanding mysteries in particle physics more than 90 years after its astrophysical discovery \citep{Zwicky1933,Zwicky1933Eng2009}. As emphasized by the 2023 P5 Report \citep{p5report2023}, cosmic survey experiments can constrain the fundamental nature of dark matter by measuring the clumping of dark matter on cosmic scales. Toward this goal, the DESI Collaboration is pursuing a Key Project to utilize this full sample of $\sim$10 million stellar spectra  \citep{cooper23} in combination with Gaia \citep{gaia21} data to characterize the distribution of matter in and around our Milky Way galaxy.  This measurement will provide new estimates of the local mass density, which are critical to derive dark matter cross section constraints in direct detection experiments. Furthermore, this measurement will serve as a first step toward testing models of dark matter particle physics directly with the DESI stellar data.

Spec-S5 will explore a large domain of dark matter models both through galaxy maps and observations of stars, with sensitivities from light to heavy particles.

The galaxy maps will explore dark matter scenarios from deviations of the power spectrum from the standard model.  For instance, a fraction of the dark matter could be non-cold, i.e. an ultra-light axion \citep{Rogers:2023ezo,Adams:2022pbo} or other light relics \citep{Dvorkin:2022jyg}, or new interactions can be present on cosmological scales\citep{Bottaro:2023wkd,2024arXiv240718252B}.  For any of these scenarios, Spec-S5 could provide a robust detection or leading constraints.

For heavier dark matter particles, Spec-S5 will greatly expand upon DESI observations by providing a sample of 50 million stellar spectra to explore the particle properties of dark matter as a secondary science goal, as recommended in the 2024 P5 report.  Following the science drivers in Chapter 4.1.6 of the 2023 P5 report, Spec-S5 will perform observations of satellite galaxies and stellar streams that will allow tests of hidden-sector dark matter models that impact structure-formation on small scales. Observations from Spec-S5 will be unique and complementary to dark matter studies with other celestial (e.g., Rubin LSST, CMB-S4) and terrestrial (e.g., direct detection, collider, and beam dump) experiments.

\begin{figure}[htbp!]
    \centering
    \includegraphics[width=0.25\textwidth]{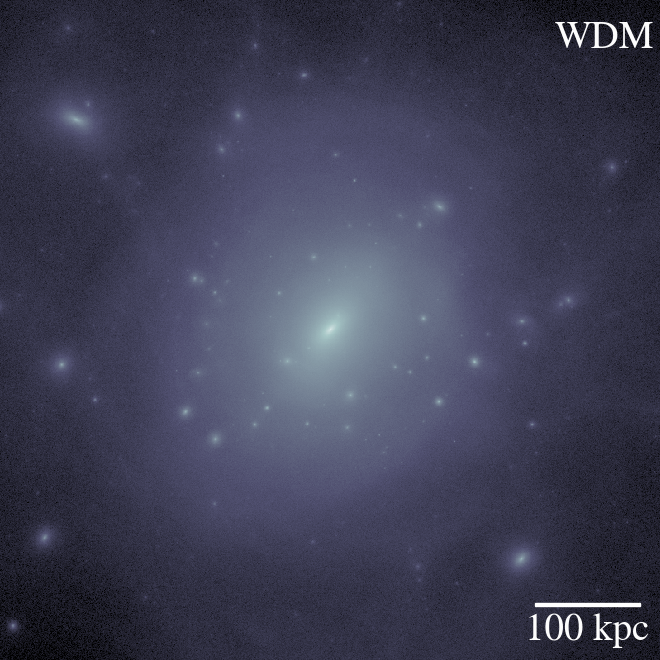}
    \includegraphics[width=0.25\textwidth]{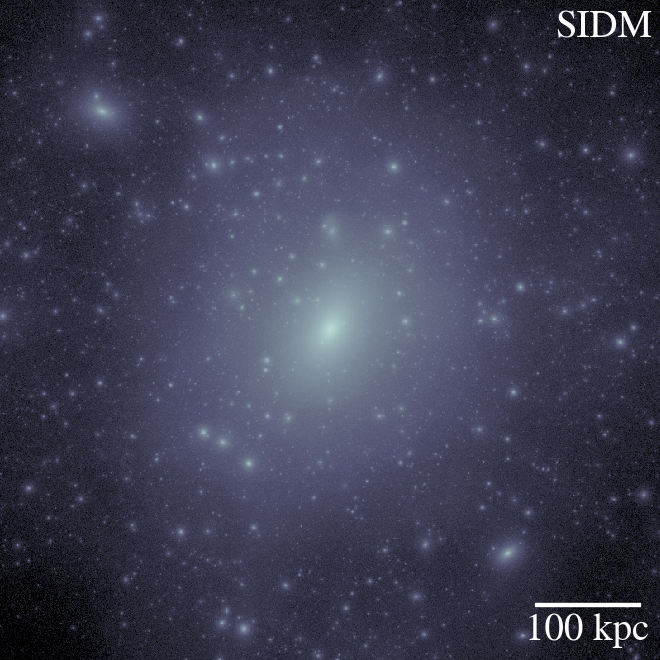}
    \includegraphics[width=0.25\textwidth]{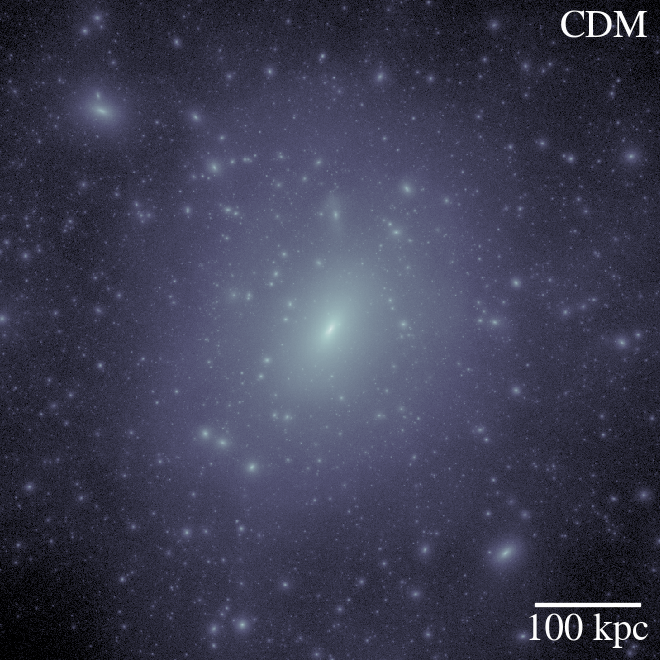}
    \includegraphics[width=7in]{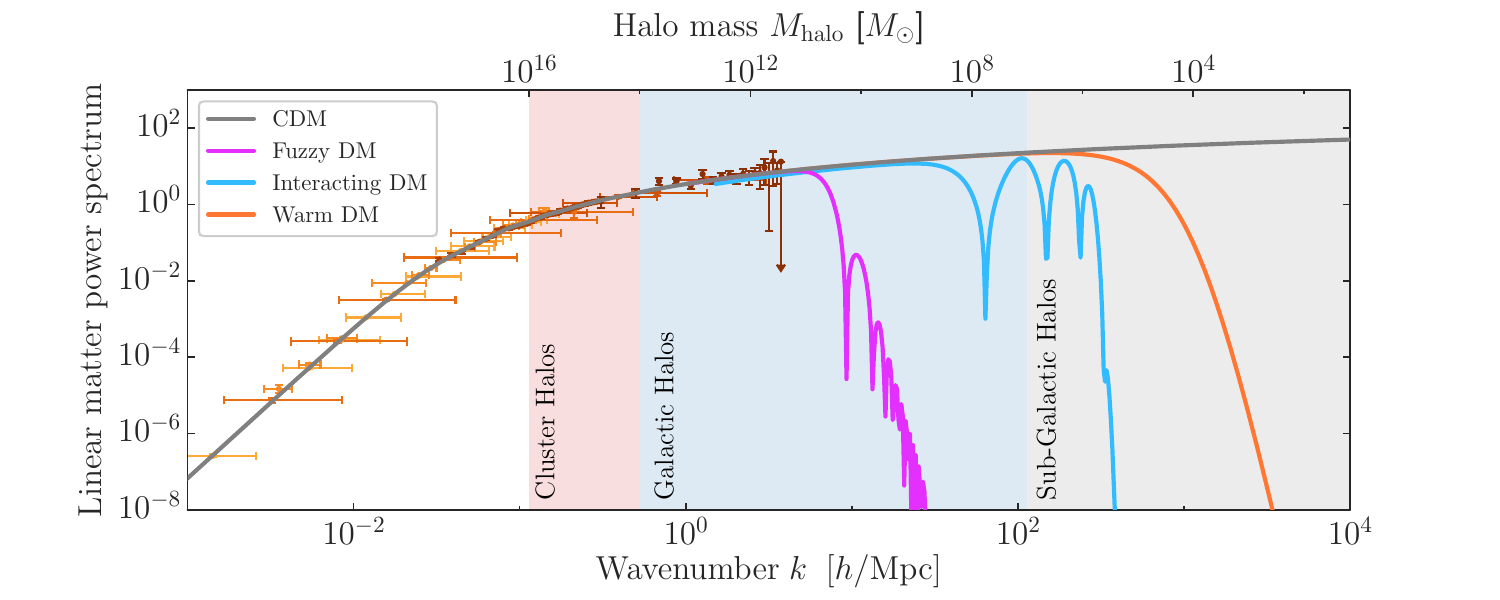}
    \caption{\label{fig:DM_summary}
    The fundamental physics of dark matter can change the abundance of dark matter substructure around the Milky Way.
    The upper panels show the distribution of dark matter around a Milky-Way-mass galaxy in warm dark matter (WDM; left), self-interacting dark matter (SIDM; middle) and cold dark matter (CDM; right) scenarios \citep[adapted from][]{BullockBoylanKolchin2017}.
    The lower panel shows the dimensionless linear matter power spectrum extrapolated to $z=0$.
    Theoretical predictions are plotted for the fiducial CDM scenario and three alternative particle physics models for dark matter (adapted from \citep{snowmassDM22}). 
    Spec-S5 is designed to explore the mass distribution of dark matter halos from the cluster scale ($\sim 10^{15}$\,M$_\odot$) to below the threshold of galaxy formation ($\sim 10^6$\,M$_\odot$).}
\end{figure}

As emphasized in the 2021 Snowmass report \citep{snowmassDM22,snowmassCF3,snowmassCF}, sampling the halo mass function at the boundary between galactic and sub-galactic scales ($k \sim 100h\,{\rm Mpc}^{-1}$) will allow simultaneous tests of the astrophysical processes and dark matter physics that govern the faint end of galaxy formation (see Figure~\ref{fig:DM_summary}).  Dwarf galaxies are extremely dark-matter-dominated systems, and Spec-S5 provides the capability to map the dark matter halo density profiles of the higher-mass dwarfs, as well as determine the kinematics of lower-mass dwarf galaxies out to their tidal radii (including new dwarf galaxies discovered by Rubin LSST). In addition, using spectroscopy of several hundred thousand stars in the Milky Way's halo, Spec-S5 will densely sample the spatially resolved kinematics of cold structures in the Milky Way halo and thus detect or constrain the existence of dark matter halos that are devoid of galaxies. 
Detailed velocity mapping of stellar streams will enable searches for disturbances from low-mass dark halos that are otherwise invisible to detection, thus probing even smaller mass scales (again, see Figure~\ref{fig:DM_summary}).
By measuring the dynamics of dwarf galaxies, stellar streams, and the Milky Way halo stellar populations, Spec-S5 can measure the mass function of dark matter halos at and below the threshold of galaxy formation, the regime where several well-motivated particle physics models of dark matter would clearly manifest directly observable signatures.

\subsection{Galaxy Formation and Evolution}

The Spec-S5 key projects will revolutionize our understanding of galaxy formation and evolution over much of cosmic history ($0 < z < 4.5$).
With spectra of more than a hundred million galaxies, this huge, new dataset will enable a detailed look into the processes that govern the assembly of galaxies over the last 12 billion years of cosmic history through measurements of stellar mass, star formation rates, metallicity, and kinematics, which will allow us to trace how these characteristics evolve over cosmic time and how they relate to the large-scale structure of the universe. Spec-S5's unique ability to measure such a vast number of galaxies at 
the epoch of ``cosmic noon'' (i.e., $z\sim2-3$) will be particularly valuable to understand the peak period of star formation, particularly the role of environmental factors that connect to the large-scale structure of the universe. These measurements will be key to identifying and correcting any astrophysical biases in the use of these galaxies as cosmological tracers. 

Spec-S5's all-sky coverage guarantees access to the entire Rubin LSST and Euclid footprints, as well as complete overlap with all other ongoing and planned multiwavelength, multimessenger and transient surveys. 
This will enable new studies of the transient universe, and provide the potential to uncover new, larger samples of rare populations of galaxies, such as those undergoing major mergers or quenching events, very young and metal poor systems, and dwarf AGN. These (relatively rare) phenomena are crucial to understanding the mechanisms that drive galaxy evolution, including feedback from active galactic nuclei (AGN) and supernovae, and the role of dark matter in shaping the growth of galaxies. The broad wavelength coverage of Spec-S5 will also provide new insights into the chemical enrichment of the intergalactic medium through measurements of outflows from galaxies and the detection of intergalactic absorption systems. The National Academies' Astro2020 Decadal Report \citep{Astro2020} emphasized the importance of understanding the physics of galaxy evolution across cosmic time, the importance of mapping large-scale structure to study the relationship between galaxies and their environment, and the key role of massively multiplexed spectroscopy. 

\begin{turnpage}
\begin{table}[ht]
\small
\caption{Yield of Stellar and Galaxy Velocities from Recent, Current and Planned Surveys}
\hspace{-0.2in}
\begin{tabular}{|l|l|c|r|r|r|r|r|r|} 
\hline
Instrument & Telescope & Nights & Star & \multicolumn{2}{c|}{Galaxy Redshifts} & Area & Operation & Ref \\
\cline{5-6}
 & & / year& RVs & $(z<2.1)$ & $(z>2.1)$ & [deg$^2$] & [years] & \\
\hline
SDSS/BOSS/eBOSS & Sloan 2.5m & all & 1.0M & 3.5M & 0.34M & 10,000 & 1999-2019 & \citep{2022ApJS..259...35A} \\
SDSS/APOGEE & Sloan 2.5m \textbf{(a)} & all & 0.73M & -- & -- & all-sky & 2011-2019 & \citep{2022ApJS..259...35A} \\
\cline{5-6}
SDSS-V & Sloan 2.5m \textbf{(a)} & all & 6.0M & \multicolumn{2}{c|}{0.4M} & all-sky & 2020-2027 & \citep{kollmeier2017} \\
\cline{5-6}
LAMOST & 3.6-4.9m & all & 11.1M & 0.27M & 0.02M & 15,000 & 2012- & \citep{LAMOST_Overview2022,LAMOST_QSO2023} \\
Gaia & 0.72m$^2$ space & all & 7.2M & -- & -- & all-sky & 2014-16 & \citep{GaiaDR2} \textbf{(b)} \\
HETDEX & HET 9.2m & 60 & -- & -- & 1M & 540 & 2014- & \citep{HETDEX_SurveyDesign} \\
DESI & Mayall 4m & all & 20M & 40M & 0.8M & 14,000 & 2021-26 & \citep{DESI2016_Part1} \\
Euclid & 1.2m space & all & -- & 35M & -- & 15,000 & 2024- & \citep{laureijs2011eucliddefinitionstudyreport} \\
Sumire/PFS & Subaru 8.2m & 20 & 1M & 3.5M & 0.5M & 800/1400 & 2025- & \citep{PFS2014_Science} \\
4MOST & VISTA 4m & all & 16M &  12.3M & 0.1M & 15,000 & 2025- & \citep{4MOST_SurveyStrategy2019} \textbf{(c)} \\
Roman & 2.4m space & all & -- &  12M & 1M & 2,000 & 2027- & \citep{RomanHLS} \\
\hline
DESI-ext& Mayall 4m & all &  & 20M & 0.2M  & 17,000 & 2025-28 &  \\
MUST & New 6.5m & all & -- & 126M & 40M & 11,000 & 2030's & \citep{MUST2024} \\
MSE & New 11.25m & all & 10M &  7M & 0.6M & 3200/80 & 2030's & \citep{MSE_Science}\textbf{(d)}\\
\cline{5-6}
WST & New 12m & all & 25M & \multicolumn{2}{c|}{250M} & 18,000 &  2030's & \citep{WST_Overview2024} \textbf{(e)} \\
\cline{5-6}
\hline
MegaMapper& New 6.5m & all & 150M & 76M & 112M & 18,000  & 2030's & \citep{2022arXiv220904322S,2022arXiv220903585S} {\bf (f)} \\
\hline
{\bf Spec-S5} & {\bf 2 x 6m } & {\bf all} & {\bf 150M} & {\bf 76M} & {\bf 112M} & {\bf all-sky} & {\bf 2036-} &  {\bf (g)}\\
 \hline
\end{tabular}\\
\rule{0in}{1.2em}\scriptsize
\textbf{(a)} In combination with the du Pont 2.5m. \\
\textbf{(b)} Only stars with radial velocities. \\
\textbf{(c)} See also \citet{4MOST_CosmoSurvey}.\\
\textbf{(d)} MSE plans are currently on hold. \\
\textbf{(e)} See also \citet{WST_Science2024}. \\
\textbf{(f)} Number of velocities and redshifts updated to match Spec-S5. \\
\textbf{(g)} See tables \ref{tab_targets_dark} \& \ref{tab_targets_bright}. \\
\label{tab_telescopes}
\end{table}
\end{turnpage}

\section{Spectroscopic Landscape}
\label{sec:Landscape}

The Spec-S5 science program requires a significant advance beyond prior instruments including those still under construction, as summarized in Table~\ref{tab_telescopes}.  SDSS/BOSS, SDSS/eBOSS and HETDEX were Stage 3 experiments operating in the 2010s that mapped a total of approximately 5 million galaxies and quasars.  DESI, the first Stage 4 experiment to come online, began operations in 2021 and has already measured redshifts for nearly 40 million extragalactic sources.  Euclid has begun operations and Sumire/PFS, 4MOST, and Roman are the other Stage 4 experiments coming online in the next several years, although none of these platforms are markedly more capable than DESI.  Most notably, the combination of all other experiments in the coming decade will only map approximately 3 million objects at redshifts $z>2.1$; DESI-2 is expected to double this.  However, none of these instruments are capable of conducting the Spec-S5 experiment in less than 50 years of dedicated operations.

In the longer term, there are four new telescopes in a planning stage that could potentially host instrumentation capable of conducting the Spec-S5 program.  The MUltiplexed Survey Telescope (MUST) would be a new 6.5-meter telescope sited on Saishiteng Mountain in China.  The Maunakea Spectroscopic Explorer \citep[MSE;][]{MSE_Science} would be a new 11.25-meter telescope sited in the CFHT Telescope dome at Mauna Kea, Hawai`i.  The Wide-Field Spectroscopic Telescope \citep[WST;][]{WST_Overview2024} would be a new 10-meter-class telescope sited at Cerro Paranal, Chile.  MegaMapper \citep{schlegel2022megamapperstage5spectroscopicinstrument} would be a new 6.5-meter telescope sited in Chile.  Spec-S5 aims to be more timely with its re-use of existing telescope platforms, its all-sky coverage, and its focus on a dedicated experimental approach.  

The Rubin Observatory photometric redshift sample of galaxies \citep{Graham2018} and SPHEREx \citep{10115793} are not included in this comparison since they have relatively poorer radial resolution for mapping large-scale structure.

\section{Survey Reference Design}
\label{sec:Survey}

\subsection{Footprint}
\label{sec:Footprint}

\begin{figure}[htbp]
    \centering
    \includegraphics[width=0.9\textwidth]{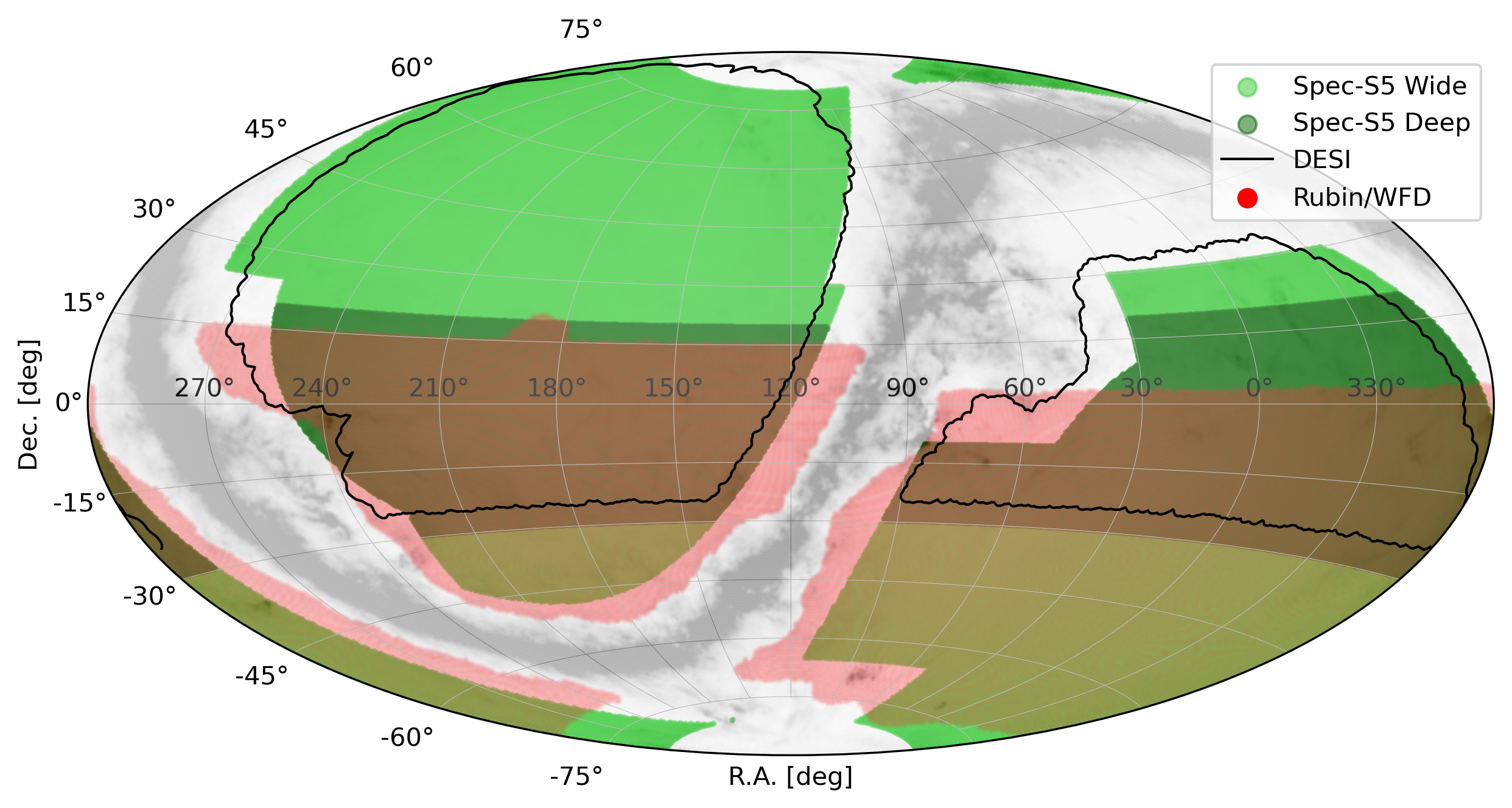}
    \caption{\label{fig:footprint}
    The footprints of the Wide-Area Cosmology Survey (light green) and the Deep Survey of high-redshift galaxies (dark green) are shown, compared with the nominal footprints of the ongoing DESI and DESI-Extension spectroscopic surveys (black line) and the Rubin Legacy Survey of Space and Time (pink).}
\end{figure}

The Spec-S5 reference survey footprints are shown in Figure~\ref{fig:footprint}. The Wide-Area Cosmology Survey will cover 25{,}000 square degrees, defined as the full-sky, extragalactic footprint that has low Galactic extinction ($E(B-V) < 0.1$ mag). Coverage of such an unprecedentedly large footprint will be achieved by conducting the experiment from two distinct sites in different hemispheres. A redshift survey of 110 million galaxies and quasars over the Spec-S5 footprint will facilitate dark energy analyses that surpass those from the DESI survey of 40 million objects over 14{,}000 sq.\ deg.
A unique aspect of the Spec-S5 survey will be the inclusion of 100 million stars to accurately and precisely map Galactic extinction prior to the final selection of samples for cosmological studies. Radial velocity measurements for an additional 50 million stars will pave the way for new dynamical studies of the properties of dark matter.
Spec-S5 will also conduct a high-redshift ($z > 2$) Deep Survey in an 11{,}000 square deg footprint, notionally defined as the equatorial region ($-30^\circ < \delta < +20^\circ$) that has low Galactic extinction.  An equatorial footprint has the advantage of being observable from either hemisphere. This adds flexibility for both identifying Spec-S5 targets and for optimizing the cadence of spectroscopic follow-up from each Spec-S5 site. The exact declination limits for the Deep Survey could be shifted either north or south, to better meet the ultimate experimental goals, while preserving the total survey area.  A key goal will be to maximize overlap with Rubin LSST and CMB-S4 coverage. 

\subsection{Targets}
\label{sec:Targets}

The Spec-S5 reference surveys include both dark- and bright-time targets, as summarized in Tables~\ref{tab_targets_dark} and \ref{tab_targets_bright}.  The dark-time targets will be observed only when the seeing, sky brightness and transparency are suitable.  The bright-time targets will be observed at other times, such as when the Moon is bright.
A number of the Spec-S5 lower-redshift target samples are similar to those for DESI, but selected to higher densities and uniformity. The higher densities are facilitated by the much faster survey speed of Spec-S5, and the improved uniformity is enabled by a combination of deeper imaging and the improved Galactic extinction maps mentioned in \S\ref{sec:Footprint}.
\begin{table}[ht]
\centering
\caption{Spec-S5 dark-time targets}
\begin{tabular}{ |l|r|r|r|r|c|r|r| }
\hline
Sample & Redshift & Area & Density & Number & Imaging & Time & Total \\
 & & [deg$^{2}$] & [deg$^{-2}$] & & source & [sec] & [hrs] \\
\hline
LRGs & $0.4-1$ & 25,000 & 1400 & 35M & Legacy+Rubin &450 &4.4M \\
ELGs & $0.6-1.6$ & 25,000 & 1400 & 35M & Legacy+Rubin & 450 & 4.4M \\
QSOs & $<2.1$ & 25,000 & 250 & 6.2M & Legacy+Rubin & 450 & 0.8M \\
LyA QSOs & $>2.1$ & 25,000 & 80 & 2.0M & Legacy+Rubin & 3600 & 2.0M \\
LBGs & $2.0-4.5$ & 11,000 & 2500 & 27.5M & Rubin & 5400 & 41M \\
LAEs & $2.1-3.5$ & 11,000 & 3000 & 33M & MB+Rubin & 2700 & 25M \\
\hline
Total & & & & 138.7M & & & 77.6M \\
 \hline
\end{tabular}
\label{tab_targets_dark}
\end{table}
\begin{table}
\centering
\caption{Spec-S5 bright-time targets}
\begin{tabular}{ |l|r|r|r|r|c|r|r| }
\hline
Sample & Redshift & Area & Density & Number & Imaging & Time & Total \\
 & & [deg$^{2}$] & [deg$^{-2}$] & & source & [sec] & [hrs] \\
\hline
BGS (galaxies) & $<0.4$ & 25,000 & 1300 & 32.5M & Legacy+Rubin & 240 &2.2M \\
GETS (stars) & n/a & 25,000 & 4000 & 100M & Gaia & 120 & 3.3M \\
DMTS (stars) & n/a & 25,000 & 2000 & 50M & Legacy+Rubin & 240 & 3.3M \\
\hline
Total & & & & 182.5M & & & 8.8M \\
 \hline
\end{tabular}
\label{tab_targets_bright}
\end{table}

\subsubsection{Bright Galaxy Sample (BGS)}
The Bright Galaxy Sample (BGS) will target the brightest galaxies on the sky, including local galaxy populations to redshifts of approximately $z < 0.4$.  Spec-S5 will increase the density of these galaxies from the 1000 per sq.\ deg.\ observed by DESI to 1300 per sq.\ deg.\ over the full 25{,}000 sq.\ deg.\ footprint.

\subsubsection{Luminous Red Galaxies (LRGs)}
The Luminous Red Galaxy (LRG) sample will target the reddest, most luminous galaxies in the universe to redshifts of approximately $z < 1$.  Spec-S5 will increase the density of these galaxies from the 550 per sq.\ deg.\ observed by DESI to 1{,}400 per sq.\ deg.\ over the full 25{,}000 sq.\ deg.\ footprint.

\subsubsection{Emission Line Galaxies (ELGs)}
The Emission Line Galaxy (ELG) sample will target star-forming galaxies with strong emission lines that make them suitable for efficiently measuring redshifts.  Variants of these galaxies form the bulk of the objects mapped by DESI, Euclid, and Roman. Spec-S5 will select a larger and more homogeneous sample of ELGs, spanning a wider area than targeted by these other surveys.\\

\subsubsection{Tracer Quasars (QSOs)}
The tracer quasar (QSO) sample will target redshifts of approximately $1 < z < 2$, where quasar numbers peak, although at number densities lower than for ELGs.  Improved imaging and deeper spectroscopy will enable Spec-S5 to increase the density of these targets from the 170 per sq.\ deg.\ observed by DESI to 250 per sq.\ deg.

\subsubsection{Lyman-Alpha Quasars (LyAQs)}
The Lyman-$\alpha$ Quasar (LyAQ) sample will target quasars at redshifts $z > 2.1$ that are sufficiently luminous to allow large-scale structure to be measured from absorption by hydrogen in the Lyman-$\alpha$ Forest along the line-of-sight to each quasar.  Improved imaging and deeper spectroscopy will enable Spec-S5 to increase the density of these targets from the 60 per sq.\ deg.\ observed by DESI to 80 per sq.\ deg. The signal-to-noise ratio for BAO measurements in the Lyman-$\alpha$ Forest will be improved by the Spec-S5 effective exposure times of 1 hour, which would correspond to 2.25 hours of DESI exposure time.

\subsubsection{Lyman Break Galaxies (LBGs)}
The Lyman Break Galaxy (LBG) sample will target the most massive and luminous galaxies at $z > 2$. These galaxies are identifiable in broad-band imaging because the Lyman Break spectroscopic feature is redshifted into the optical bands. LBGs can be efficiently selected from the deep, broad-band imaging provided by the Rubin Observatory LSST.  Spec-S5 will select $u$-band drop-out galaxies to a magnitude limit of $r<24.5$ and $g$-band drop-out galaxies to a magnitude limit of $i<24.7$.  In combination, this will provide a sample of 3{,}000 galaxies per sq.\ deg.\ spanning redshifts of approximately $2.0 < z < 4.5$.
The selection of LBGs for Spec-S5 has been demonstrated using existing Subaru HyperSuprimeCam imaging, which has similar characteristics to data expected from the Rubin Observatory LSST.  Redshifts for representative samples of these targets have been obtained with DESI, in part as a pilot program for Spec-S5.

\subsubsection{Lyman Alpha Emitters (LAEs)}
The Lyman Alpha Emitter (LAE) sample will target galaxies that are less massive than the LBG population but that can have efficiently measured redshifts due to bright Lyman-$\alpha$ emission. These objects can be identified in medium-band imaging surveys where the Lyman-$\alpha$ emission enhances the flux in one of these filters as compared to broad-band filters.  A set of 7 filters spanning wavelengths of $375 < \lambda < 550$\,nm can be used to target LAEs in the redshift range $2.1 < z < 3.5$.
The selection of LAEs for Spec-S5 has been demonstrated using existing Subaru SuprimeCam imaging and the ongoing DECam Intermediate-Band Imaging Survey program discussed in \S\ref{sec:imaging}. 
Redshifts for representative samples of these targets have been obtained with DESI, in part as a pilot program for Spec-S5.

\subsubsection{Galactic Extinction Tracer Stars (GETS)}
\label{sec:extinctionstars}
Spec-S5 will improve the uniformity of large-scale-structure studies by deriving a Galactic extinction map from distant Milky Way stars that sample nearly the full column of intervening dust at high Galactic latitude.  A comparison of the observed colors of these stars to the predicted colors based upon spectroscopy results in a high-fidelity measure of extinction.
A sample of 4000 stars per sq.\ deg\ at $G<20$ mag will be selected from the {\em Gaia} catalog. Exposures will be sufficiently long to model the intrinsic colors of the observed stars from their spectral features.  This approach has been demonstrated using BOSS data \citep{schlafly11} and DESI data \citep{zhou25}, but with far fewer stars.  Spec-S5 will provide a sample of 100 million stars to construct Galactic extinction maps with an angular resolution of $\sim 1$ arcmin.

\subsubsection{Dark Matter Tracer Stars (DMTS)}
\label{sec:DMstars}
Imaging surveys have yielded large samples of stars in the Milky Way halo, including many belonging to dynamically coherent structures such as dwarf galaxies, stellar streams, and shells. Samples of known halo stars and halo substructures will greatly increase with the Rubin Observatory LSST \citep[see, e.g.,][]{Rubin_LSST_Science_Book_2009}. Spec-S5 will leverage photometric and astrometric catalogs from {\em Gaia}, Rubin LSST, and {\em Roman} to target halo stars. In particular, Spec-S5 will prioritize stars in stellar streams and dwarf galaxies in order to try to constrain the properties of dark matter.
Color-magnitude information will be the primary means of selection, supplemented by astrometric and proper-motion information from {\em Gaia}  for brighter stars and Rubin LSST/{\em Roman} for fainter stars.
A sample of 50 million stars will be targeted during the Spec-S5 bright-time program.

\subsubsection{Time Domain Studies}

In addition to mapping tracer populations, Spec-S5 will be poised to enrich a number of studies in the time domain:

\begin{itemize}
    \item \textbf{Serendipitous searches for, and follow-up of, tidal disruption events (TDEs)}: Over a 10-year baseline, Rubin is expected to identify between 10{,}000 and 80{,}000 TDEs \citep[][\href{https://docushare.lsst.org/docushare/dsweb/Get/Document-37622/TDE_cadence_note.pdf}{Rubin cadence note by van Velzen et al. 2020}]{Bricman2020}. Spec-S5 could potentially initiate high-cadence spectroscopic follow-up of TDEs to trace the evolution of the post-TDE phase, which could help to characterize star composition and SMBH-star pollution signatures \citep{Floris2024}. Spectroscopic follow-up with Spec-S5 could be crucial to help discern TDEs, supernovae, and flares from Active Galactic Nuclei (AGNs).
    
    \item \textbf{Follow-up of Supermassive Binary Black Holes (SMBBHs)}: The Square Kilometer Array Pulsar Timing Array \citep{Feng2020} is expected to discover $\sim$100 SMBBHs/yr, and Rubin \citep{Davis2024} is likely to also augment the population of known SMBBHs. However, reliably detecting SMBBHs, and assessing false-positive rates, will be a challenge that can potentially be addressed by follow-up spectroscopy with Spec-S5.
    
    \item \textbf{Changing-Look/Changing-State AGNs and AGN flares}: SDSS IV studies suggest that fewer than 1\% of AGNs are in a flaring or changing-look (CL) state \citep[e.g.][]{Ricci2023}. However, investigations with SDSS I-IV \citep{Panda2024}, SDSS V \citep{Zeltyn2024} and the DESI Early Data Release \citep{Guo2024} suggest that longer baselines are needed to test biases inherent to detecting fewer Turn-On versus Turn-Off CL-AGNs. These baselines could be achieved using spectroscopic observations spanning from the SDSS era to the Spec-S5 era. 
    
    \item \textbf{Reverberation Mapping (RM) of AGNs}: Spectroscopic RM campaigns are beginning to encompass significant populations of AGNs. For example, the SDSS V Black Hole Mapper is conducting an RM campaign of more than 1000 AGNs, spanning a wide range of redshifts ($0.1 < z < 4.5$) and bolometric luminosities (10$^{44}$ - 10$^{47.5}$ erg s$^{-1}$) to $i < 20$ \citep{Shen2024}. In addition, the 4MOST RM campaign is expected to ultimately sample of order 1000 AGNs with $0.1 < z < 2.5$ and $r < 22.5$. Over 10 years, Rubin should detect time lags for $\sim$2 million AGNs \citep{Kovacevic2022}, and will begin to populate the low-luminosity, low-mass end of the AGN census \citep{Czerny2023}. By following up Rubin candidates, Spec-S5 would have the capability to improve on existing RM studies in terms of both AGN numbers and depth.

\end{itemize}

\subsection{Imaging Survey}
\label{sec:imaging}
Spec-S5 will select targets using a combination of existing publicly-available imaging and new, dedicated imaging surveys. 

The baseline plan for selecting the high-redshift (LAE and LBG) targets for the Spec-S5 Deep Survey is to use data from the Rubin Observatory. The LBGs will be selected from the first five years of Rubin's LSST. The LAE targets will be selected by dedicating three years of the Rubin telescope, after LSST is complete, to a medium-band survey of the 11,000~deg$^2$ Spec-S5 high-redshift footprint. This imaging will span a wavelength range of 375--550\,nm to efficiently select LAEs over the redshift range $2.1 < z < 3.5$. To achieve a constant comoving number density, this imaging is required to reach $6\sigma$ depths, which vary from $\approx 1.17\times10^{-16}$~erg/s/cm$^2$ in the bluest filter to $\approx 4.3\times10^{-17}$~erg/s/cm$^2$ in the reddest medium-band filter.  
This imaging would not be required until after Spec-S5 operations have begun. 

%

If the Rubin LSST survey is delayed (thus delaying the start of a Rubin-2 medium-band survey), the LAE targets could be initially selected by combining Rubin LSST broad-band imaging with (ongoing and planned) medium-band imaging surveys over the high-redshift footprint.  
The DESI-2 project has already initiated a medium-band imaging program over an initial area of 1000~deg$^2$ using Blanco/DECam (the Intermediate Band Imaging Survey, or IBIS, which commenced in May 2024), and intends to expand this footprint to as much as 5000~deg$^2$ using a combination of the 
CFHT/MegaCam, Blanco/DECam and Subaru/HSC cameras. 
The bluest filters (MB0 and MB1; see Table~\ref{tab_imaging}) would be observed with CFHT and filters MB2--MB5 would be observed with DECam. HSC would provide a deeper, higher-density survey over 1{,}000~deg$^2$ using a single mosaiced filter that is in production. In this scenario, any remaining portion of the 11{,}000~deg$^2$ of the medium-band imaging would be conducted with the Rubin Observatory, and could be completed in fewer than 3-years of observing during dark time.  

The imaging surveys will be conducted using a ``dynamic’’ observing strategy, designed to result in a survey of uniform depth in the minimum amount of observing time \citep[e.g][]{burleigh20}. In this approach, exposure times for each image are determined on-the-fly given the observing conditions and position of the field. This approach is currently successfully employed by DESI and DECam, and would be the optimal approach for future Rubin surveys.

The remaining targets for the reference survey (i.e., the Wide-Area Cosmology Survey targets) will be selected from existing broad-band imaging and astrometry. 
The Legacy Surveys imaging that was used to select DESI targets will be adequate to select the LRG, ELG, QSO, LyAQ, and BGS targets to the required depth. 
The DMTS (\S\ref{sec:DMstars}) target selection will combine Legacy Surveys imaging with {\em Gaia} proper motions.
Rubin will be imaging in the same broad bands as the Legacy Surveys and will ultimately supersede the Legacy Surveys within the Rubin footprint, as it is expected to achieve greater S/N and uniformity after the first 2 years of observations.  
The GETS (\S\ref{sec:extinctionstars}) will be selected from {\em Gaia} Data Release 4, which is expected to be available in 2026 with sufficient information to easily select the appropriate stars over the full sky to a magnitude limit of $G<20$.  Color information (and, where possible, DESI spectroscopy) will be used to limit the selection to \mbox{A-,} F- and G-type stars, which have been demonstrated to most accurately measure extinction.

\begin{table}
\centering
\caption{Medium-band imaging options}
\begin{tabular}{ |l||r|r|r|r|r|r|r| }
\hline
Filter           &     MB0   &     MB1   &     MB2   &     MB3   &     MB4   &     MB5   &     MB6     \\
\hline
Cut-on $\lambda$ [Ang] & 3750 & 4006 & 4265 & 4519 & 4779 & 5036 & 5297 \\
Cut-off $\lambda$ [Ang] & 4006 & 4265 & 4519 & 4779 & 5036 & 5297 & 5507 \\
Line flux [$10^{-16}$~erg/s/cm$^2$]        &    1.17 &     1.01  &    0.86  &     0.76  &     0.65  &     0.56  &     0.43    \\
N(LAE) deg$^{-2}$    &      430  &      431  &      439  &      430  &      426  &      418  &      410    \\
\hline
CFHT nights, 2000 deg$^2$ &     63  &     82  &        -  &        -  &        -  &        -  &        -    \\
DECam nights, 2000 deg$^2$ &        -  &   143  &     129  &     140  &    162   &     197  &     250    \\
HSC nights, 2000 deg$^2$       &        -  &      11  &      10  &      11  &      13  &        -  &        -    \\
Rubin nights, 9000 deg$^2$     &      18  &       20  &       21  &       25  &       31  &       38  &       49    \\
 \hline
\end{tabular}
\label{tab_imaging}
\end{table}

\subsection{Survey Plan and Operations}

Spec-S5 will conduct a highly uniform spectroscopic survey over its first 6 years of operations, with the expectation that there will be an extended operations phase informed by developments in the field. 

Spec-S5 will begin observations with a six-month Survey Validation phase at each site.  The final refinements to the target selection algorithms will be made during Survey Validation to ensure that high spectroscopic completeness is achieved within the requested exposure times.  During this phase, the visible portion of the GETS program will be observed, with 60M of those stars observed at the first site to begin operations and the remaining 40M observed on the second site.  Those stars are deemed bright-time targets, but can be observed even more quickly during dark-time, requiring only 3 months of the Survey Validation time at the first site and 2 months at the second site spaced throughout that six month period.

 Spec-S5 will employ a dynamical observing mode, similar to that which has been implemented for DESI operations. The ``effective exposure time'' (i.e., the exposure time corresponding to a defined standard set of observational conditions) will be tracked in real time during each exposure using an exposure time calculator, with each exposure ending when it reaches a specified signal-to-noise ratio limit for a fiducial target.  The dark time program will consist of one set of tiles for short (450 sec effective) exposures containing the LRG, ELG, and tracer QSO targets.  The longer (900 sec effective) dark time tiles will contain the LyA QSO, LBG and LAE targets, with those targets receiving 4, 6 and 3 exposures respectively.  The bright-time tiles will observe BGS and DMTS to a uniform 240 sec (effective) depth.  Observing will automatically switch between the dark-time and bright-time tiles depending on conditions.  This approach is nearly identical to the observing strategy for DESI, with exposure times scaled from DESI pilot observations.  Targets with several exposures will be observed on different fibers on different tiles, a strategy that proved successful for the faintest object spectroscopy using DESI.

\subsection{Survey Operations and Data Processing}
The dual-site operation for Spec-S5, combined with a 5 to 10-fold increase in data throughput compared to current surveys like DESI, requires a sophisticated operational model.
The system must process approximately 500,000 spectra nightly within 12 hours to enable dynamic targeting decisions, particularly for high-redshift quasars. Leveraging modern computing architectures and parallel processing techniques, the pipeline will ensure rapid data reduction, calibration, and redshift determination while maintaining high-quality standards. Data storage is estimated at 6 petabytes for the entire survey, with computational power expected to grow sufficiently by the survey's start to handle the increased data volume.

Real-time monitoring systems will track observing conditions and adjust integration times dynamically, while a centralized filesystem will manage target status across both hemispheres. The survey strategy adopts a ``depth-first" approach, completing observations in contiguous regions before moving to new areas, similar to DESI, to maximize scientific output. Regular calibration observations at both sites will ensure precise flux calibration, critical for Lyman-$\alpha$ forest analysis, while the dual-site operation provides redundancy and cross-checking opportunities to mitigate systematic effects. Daily quality assurance processes, initially involving manual oversight, will ensure data integrity and survey progress.

The data processing pipeline will build on DESI's proven architecture, incorporating improvements to handle the increased scale and complexity of Spec-S5. Key components include pre-processing, spectral extraction, sky subtraction, flux calibration, and redshift measurement. Efficient spare fiber management will be crucial, with a formal process to evaluate and prioritize programs aligned with key science goals. Software development will follow an open-source model, hosted on GitHub, with a modular architecture and rigorous version control to ensure compatibility and stability. Operational flexibility will allow the system to adapt to disruptions, ensuring consistent data quality and survey progress across both hemispheres.

In conclusion, Spec-S5's operational model and data processing system are designed to meet the challenges of its ambitious survey program. By building on DESI's successes and incorporating new capabilities, the system will support Spec-S5's goal of mapping the universe with unprecedented precision. The dual-site operation, while complex, offers significant advantages in efficiency and systematic error mitigation, ensuring the survey's scientific success.

\section{Instrument Reference Design}
\label{sec:Instrument}


\begin{figure}[thb]
    \centering
    \includegraphics[width=4in]{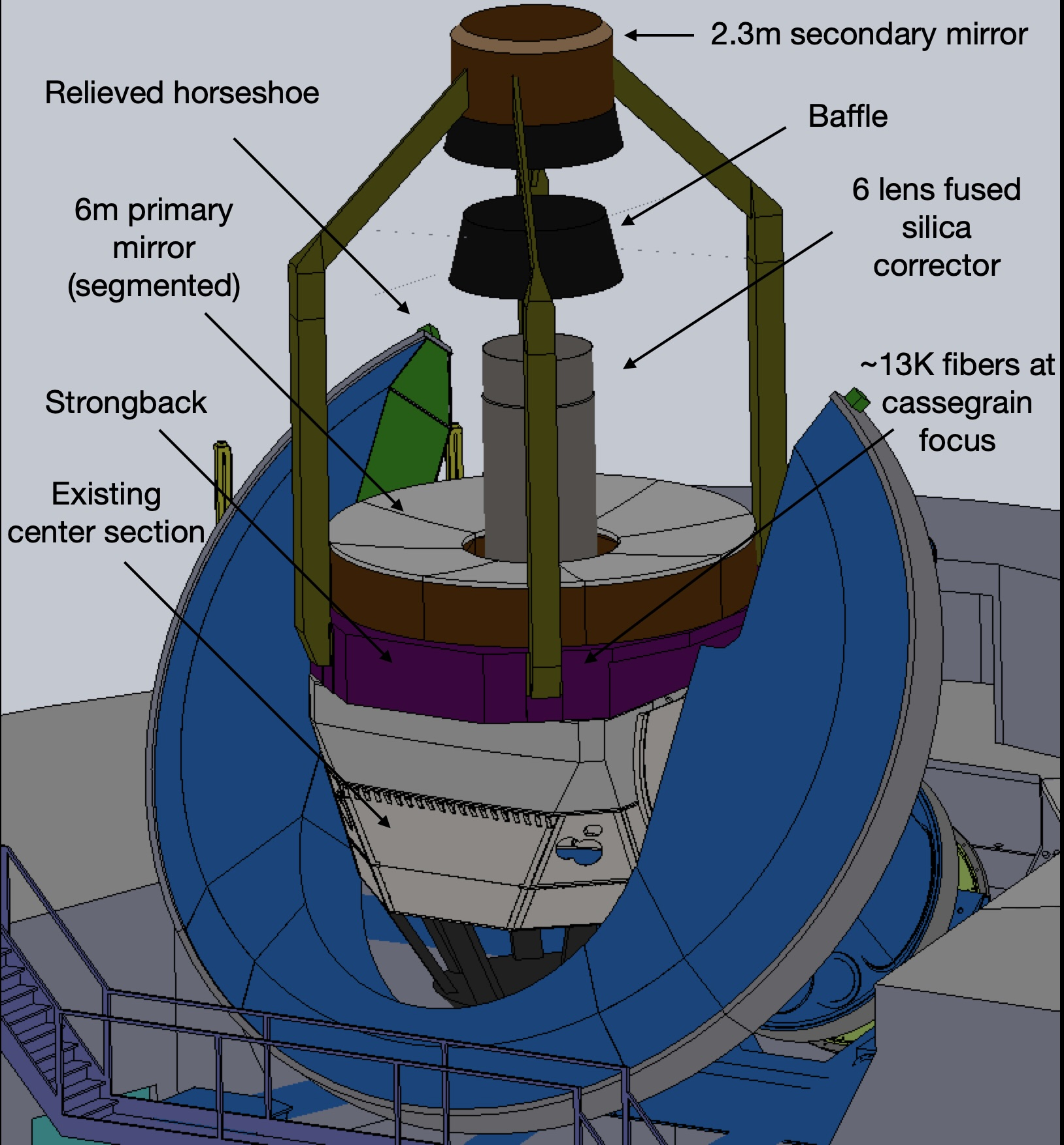}
    \caption{Conceptual design of the telescope modifications. The modification of the blue horseshoe mount provides space for a new, segmented primary mirror (M1). The secondary mirror is supported above the primary with four legs and directs light into the corrector and then the focal plane instrumentation at Cassegrain.}
    \label{fig:mount}
\end{figure}

The reference design for Spec-S5 consists of two telescopes, one in each hemisphere, to provide the necessary science reach while also providing all-sky coverage. Each 6.0-m telescope will provide a wide-field (2.2 deg diameter) focal plane located under the primary mirror at the Cassegrain focus of the telescope. The focal plane will be sampled by a highly-multiplexed fiber-fed (12,852 fibers) instrument, with the fibers feeding 23 spectrographs that are duplicates of the current DESI spectrographs.  Additional gains are made possible with improved blue throughput and lower detector read noise. Spec-S5 represents a $>12\times$ gain on the current DESI spectroscopic instrument, and is thus capable of efficiently surveying the necessary samples of faint, high-redshift galaxies and deliver unique primordial cosmological constraints. 


\subsection{Instrument Platform}

Spec-S5 will make use of two existing NSF telescope facilities, the Nicholas U.\ Mayall Telescope at Kitt Peak, Arizona, and the Victor M.\ Blanco Telescope at Cerro Tololo, Chile. Both platforms currently host successful DOE experiments and we are working with the appropriate agencies to ensure that they are capable of supporting the Spec-S5 instrument. The instrumentation and upgrades at each telescope will include: a 6-m diameter circular primary mirror (M1); a 2.3-m secondary mirror (M2); an optical corrector with six spherical lenses; a focal plane with 12,852 robotic fiber positioners, and a fiber system feeding 23 three-channel spectrographs.  The control systems and data systems would build directly upon the DESI project heritage.

These two telescopes are suitable platforms for Spec-S5 owing to their over-sized equatorial mount structures that accommodate a moving mass of 375 tons.  The existing 4-m primary mirrors and their cells have a mass of 30 tons and would be replaced with modern, 6-m mirrors and cells with less mass and superior thermal properties.  The telescope mount would be modified by widening the opening of the ``U" of the horseshoe above the declination bearings and the attachment flanges of the elliptical tubes.  The non-load bearing tips of the horseshoe will be notched as shown in Figure \ref{fig:mount}, with the resulting open area closed out by installing steel plates.  An engineered solution will ensure that the horseshoe remains structurally sound after the modification and that the hour angle journal bearing and declination bearings perform as they currently do for the duration of the project. No other major structural modifications are needed for either facility.

\subsubsection{Platform alternative analysis}
To arrive at the reference design, we investigated three different platform options for the Spec-S5 instrument: 
\begin{itemize}
    \item[(A)] Refurbish the Mayall and Blanco 4-m telescopes with $\approx$13,000 fibers each. 
    \item[(B)] Construct a new (single) 6.5-m telescope equipped with $\approx$26,000 fibers. 
    \item[(C)] Upgrade the Mayall and Blanco with 6-m mirrors and $\approx$13,000 fibers each.
\end{itemize}

Option A, involved upgrading the focal plane of the DESI instrument (on the KPNO Mayall 4-m telescope) to $\approx$13,000 fibers and then duplicating this at the Blanco 4-m telescope. This option had the benefits of using existing sites and the ability to be accurately costed based upon the DESI project heritage. However, this approach required many years of operations to accomplish the science due to the 4-m aperture of the telescopes. Option B \citep[the ``MegaMapper'' option;][]{Megamapper2022a,Megamapper2022b} had the benefit of reaching the primordial cosmology science goals of Spec-S5 in 6 years, but required the construction of a new, large facility with less certain budget and schedule (cf. Rubin). This design also required lenses larger than any previously manufactured, thus adding risk to the project. By virtue of being a single telescope in one hemisphere, it also resulted in a cosmological footprint limited to $\approx 60\%$ of the sky and lacked the all-sky capability needed for the best constraints on the low-order modes in the galaxy power spectrum and for comprehensive studies of the Milky Way. 


Engineering feasibility studies resulted in Option C, an upgrade of option A where the primary mirror size in both telescopes increases from 4-m to 6-m diameter, as the Spec-S5 reference design.  It is a unique and fortuitous discovery that these existing NSF telescope platforms can accommodate 6-m diameter mirrors with only modest modifications to the telescope mechanical structures.  The reference design of the instrument has a 3.8 sq.\ deg.\ focal plane with 12,852 fibers at each site.  This makes the telescope optics comparable to previously-built optics, and reduces the number of spectrographs required on each site to a number (23) that can be accommodated in each enclosure. This baseline preserves the full multiplex of the MegaMapper variant, with the additional benefit of full-sky coverage that enhances many of the science cases (i.e., dark energy, dark matter, Galactic extinction, and galaxy evolution).  This reference design has the lowest lifecycle costs of any of the other options available, and the use of existing and well-maintained facilities results in this being the fastest and most cost-effective path toward achieving the scientific goals. 

The northern site, the Mayall 4-m telescope at the Kitt Peak National Observatory (KPNO) near Tucson, Arizona, currently hosts the DESI instrument and is dedicated to conducting the DESI survey.  The telescope is operated by the NSF NOIRLab.  Telescope operations are currently fully funded by the DOE HEP to conduct the DESI survey.

The southern site, the Blanco 4-m telescope at the Cerro Tololo Inter-American Observatory (CTIO) near La Serena, Chile, currently hosts the DECam instrument, which conducted the DOE HEP-supported Dark Energy Survey from 2012--2019.  The telescope is currently funded by the NSF and operated by the NSF's NOIRLab for conducting PI-led observations.

We propose to utilize each of these telescopes as the two platforms for Spec-S5, modifying the telescopes as needed to accommodate the Spec-S5 instrument and to perform the dedicated Spec-S5 survey.
NSF NOIRLab has indicated in a letter that they are interested in pursuing this approach for Spec-S5.

Prior to CD-1, the DESI project conducted a technical analysis of the 20 comparable 4-m-class telescopes and sites to determine their viability for hosting a large, highly-multiplexed spectroscopic instrument; this study is still relevant today.
Only a handful of these telescopes are capable of hosting the Spec-S5 instrument due to size and weight constraints. Of these, the Mayall and Blanco telescopes are the two existing telescopes that are best-suited to the all-sky Spec-S5 program. The CFHT on Mauna Kea, Hawai`i is pursuing a long-term replacement of that telescope and is not expected to be available for Spec-S5.  The AAT in Siding Spring, Australia typically experiences fewer clear nights per year with poorer seeing than the alternatives.

We have excellent long-term data for evaluating the two sites.  The night-time sky brightness, atmospheric transmission and atmospheric transparency (cloudiness) distributions between the two sites are nearly identical.  We characterize these sites as the number of ``effective hours,'' defined as the observing time equivalent to observing during nominal, photometric, dark-time conditions.  These nominal conditions are defined as a sky brightness of 21.07 mag/arcsec$^2$, seeing of 1.10 arcsec FWHM in $r$-band, and 100\% transparency (i.e., no clouds).  Survey speed is defined as
\begin{equation}
    \rm{Speed} = f_{\rm fiber}^2 * T^2 / S
\end{equation}
where $f_{\rm fiber}$ is the fraction of a target's light being captured by the fiber in nominal seeing, $T$ is the relative atmospheric transparency, and $S$ is the sky brightness relative to a nominal 21.07 mag/arcsec$^2$.  At both the Mayall and Blanco telescopes, there are 650 effective hours of observing each year when it is appropriate to observe dark-time targets (Speed $>$ 0.4).  In addition, there are 65 effective hours in worse conditions (Speed $<$ 0.4) appropriate for observing bright-time targets.  Any given year will typically vary by 10\% from this average.  The Spec-S5 six-year survey will have 100M fiber-hours of effective time in dark-time observing and 10M fiber-hours of effective time in bright-time observing.

\subsection{Telescope Optics}

The 6-m primary mirror will be fabricated from a low-expansion glass such as ULE\textcircled{R} or Zerodur\textcircled{R}. This mirror could either be constructed as a single surface or a segmented annulus with individual maximum dimensions of approximately 2-m.  These mirrors will sit on a set of active actuators to phase the segments, maintain figure and focus based on thermal changes, and balance the load as the telescope is slewed, based upon similar meniscus mirrors for the NSF WIYN, NSF SOAR and NSF Gemini telescopes.

\begin{figure}[thb]
    \centering
    \includegraphics[width=4in]{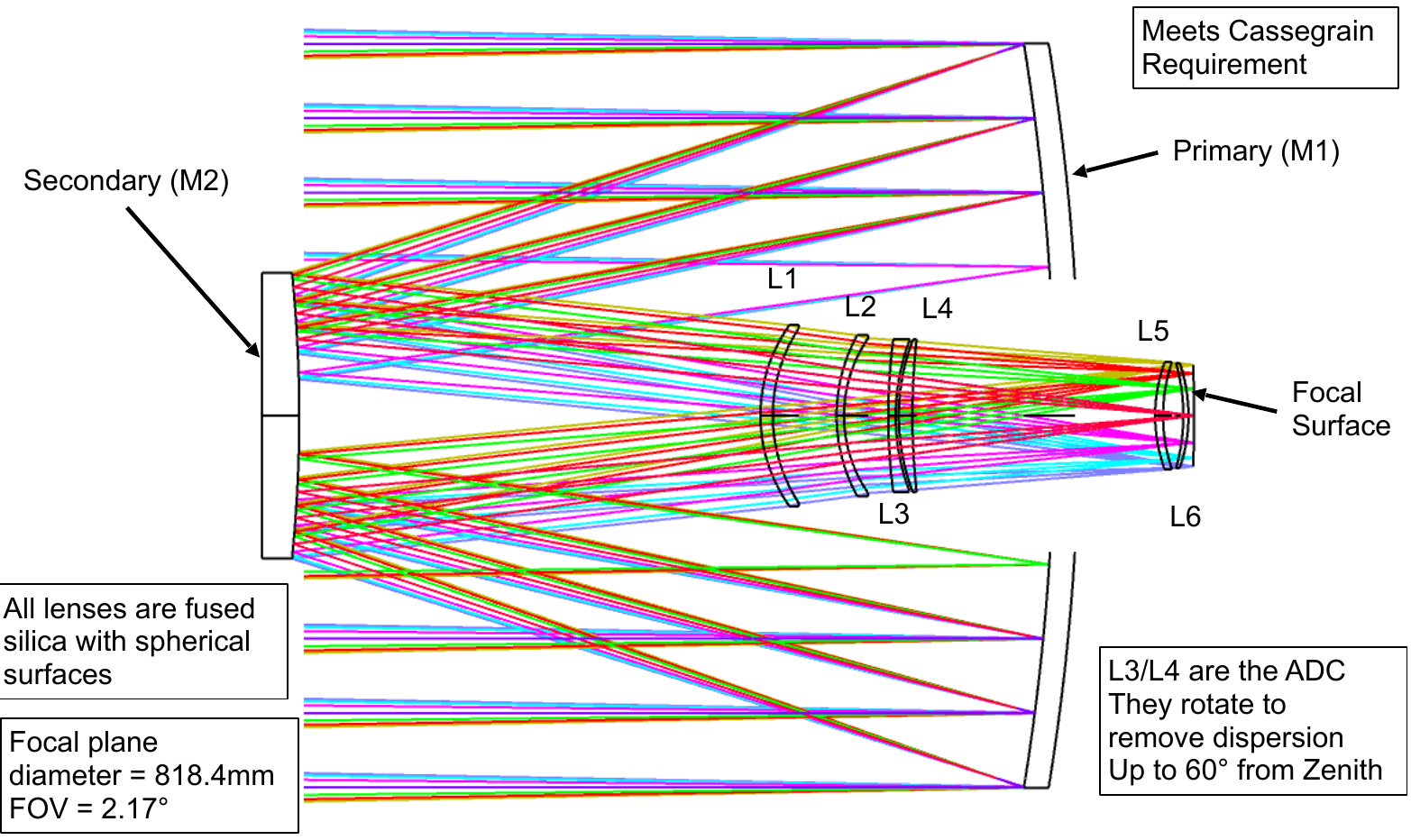}
    \caption{Optical design of the telescope mirrors and corrector.}
    \label{fig:optics}
\end{figure}

A trade study has been performed on the optical solutions to deliver a corrected field of view of at least 2 deg diameter with a physical diameter of 0.82-m located at the Cassegrain focus.  The Cassegrain can be balanced on the telescope mount, and has the additional benefit of requiring a shorter optical fiber run from the focal plane to the spectrographs (which improves the throughput, especially at shorter wavelengths).  These trade studies included a 6.5-m stadium mirror (reducing one dimension to fit within the existing mechanical mount constraints) and 6-m primary mirrors with a range of optical speeds that match the Spec-S5 spectrographs.  The optical design is shown in Figure~\ref{fig:optics} and the scale of these lenses are shown in Table \ref{tab_optics}.

\begin{figure}[htbp!]
    \centering
    \includegraphics[width=0.48\textwidth]{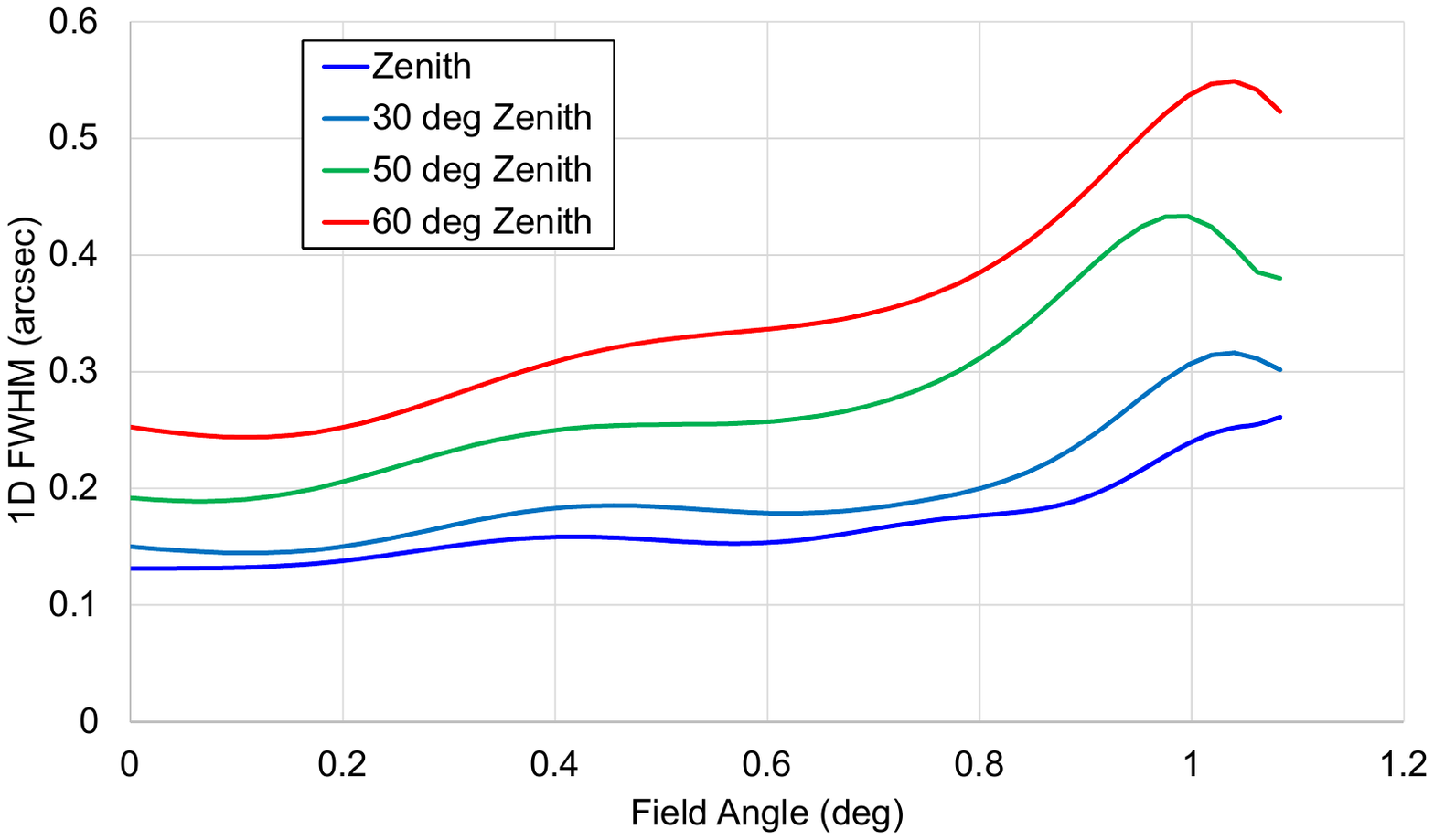}
    \includegraphics[width=0.48\textwidth]{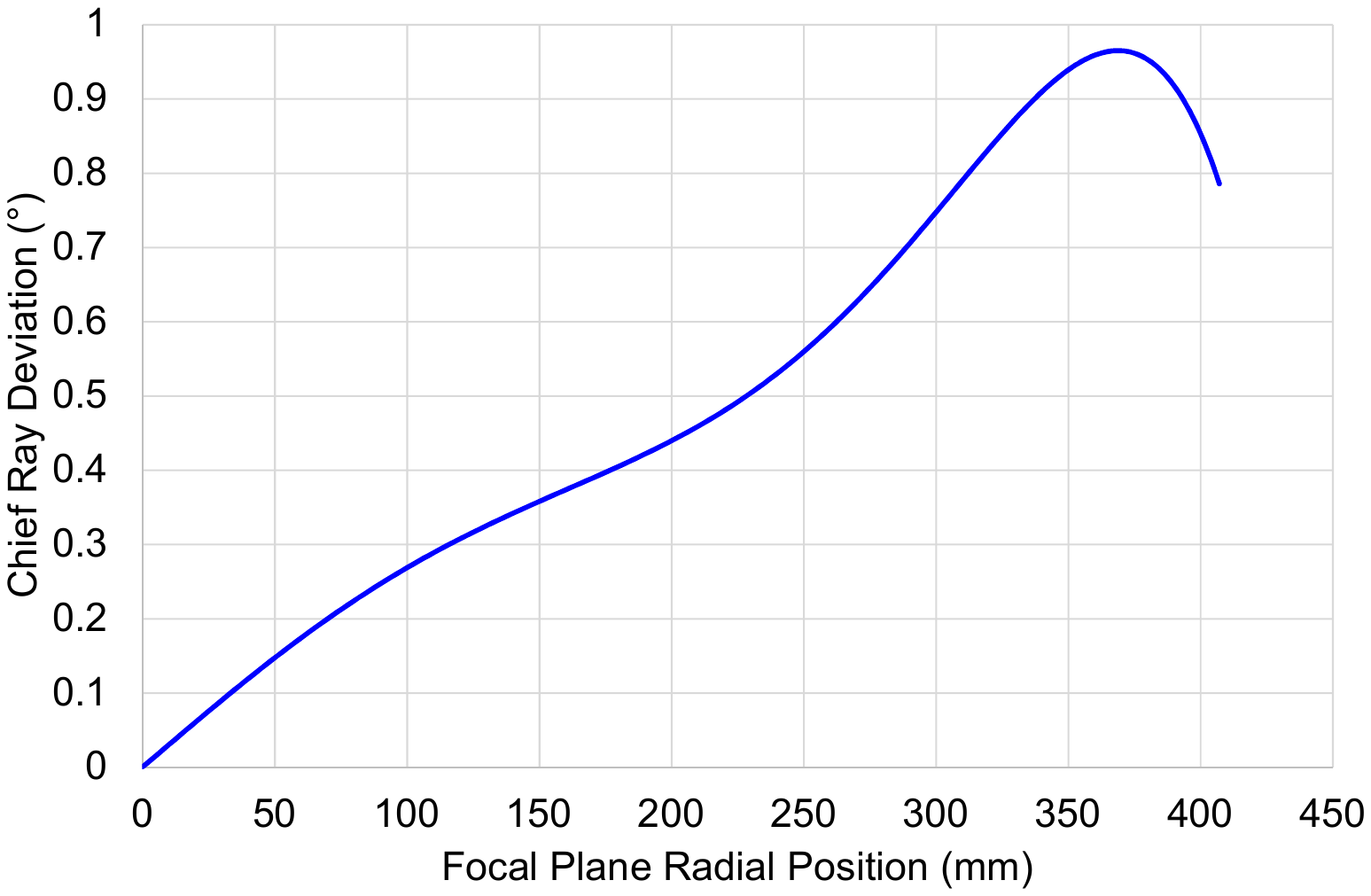}
    \caption{The reference design for the Spec-S5 optical system meets both the average and maximum blur specification for Zenith angles from 0$^\circ$ to 60$^\circ$. The maximum chief ray deviation is 0.97 deg and the focal ratio incident on the focal plane varies by $\leq$1\% in comparison to 8.9\% variation for the DESI focal plane. \label{fig:blur}}
\end{figure}

The preferred solution is the 6.0-m circular primary mirror with a 2.30-m secondary mirror and 6-element corrector (Table \ref{tab_optics}). A study is underway to understand the optomechanical requirements of M1 and M2. The corrector lenses range in diameter from 0.85-m to 1.46-m, which are smaller than the largest LSST camera lens (1.55-m).  Notably, all of the lenses are spherical surfaces, which simplifies their manufacture relative to LSST and DESI which each had several aspherical surfaces.  The corrector includes two counter-rotating lenses that act an atmospheric dispersion corrector (ADC), as was done for DESI.  The focal plane is a 0.82-m diameter surface situated under the primary mirror, allowing easy accessibility.  The optical speed at the focal plane is $f/3.6$ with only a few percent variation, which is perfectly-matched to the Spec-S5 spectrographs. The maximum chief ray deviation is 0.97 deg and Zenith angles from 0° to 60° meet both the average and maximum blur specification as shown in Figure~\ref{fig:blur}.
\begin{table}
\centering
\caption{Reference Optical design}
\begin{tabular}{ |l||r|r|c|c| } 
\hline
Optic     & Diameter  & Mass & Material  & Surface\\
   & [m] & [kg]  &  & \\
\hline
M1        & 6.00         & -        & Mirror & Conic\\
M2        &  2.30         & -       & Mirror & Conic\\
L1        & 1.46         & 351.9    & Fused Silica & Spherical\\
L2        & 1.30         & 231.5    & Fused Silica & Spherical\\
L3 (ADC1) & 1.23         & 256.8    & Fused Silica & Spherical\\
L4 (ADC2) & 1.23         & 187.6    & Fused Silica & Spherical\\
L5        & 0.87         & 98.3     & Fused Silica & Spherical\\
L6        & 0.85         & 43.6     & Fused Silica & Spherical\\
Focal Plane & 0.82         & -      & - & Asphere\\
\hline
\end{tabular}
\label{tab_optics}
\end{table}

\subsection{Focal Plane}

\begin{figure}[th]
    \centering
    \includegraphics[width=6.5in,trim={2.5cm 10cm 6cm 1.5cm},clip]{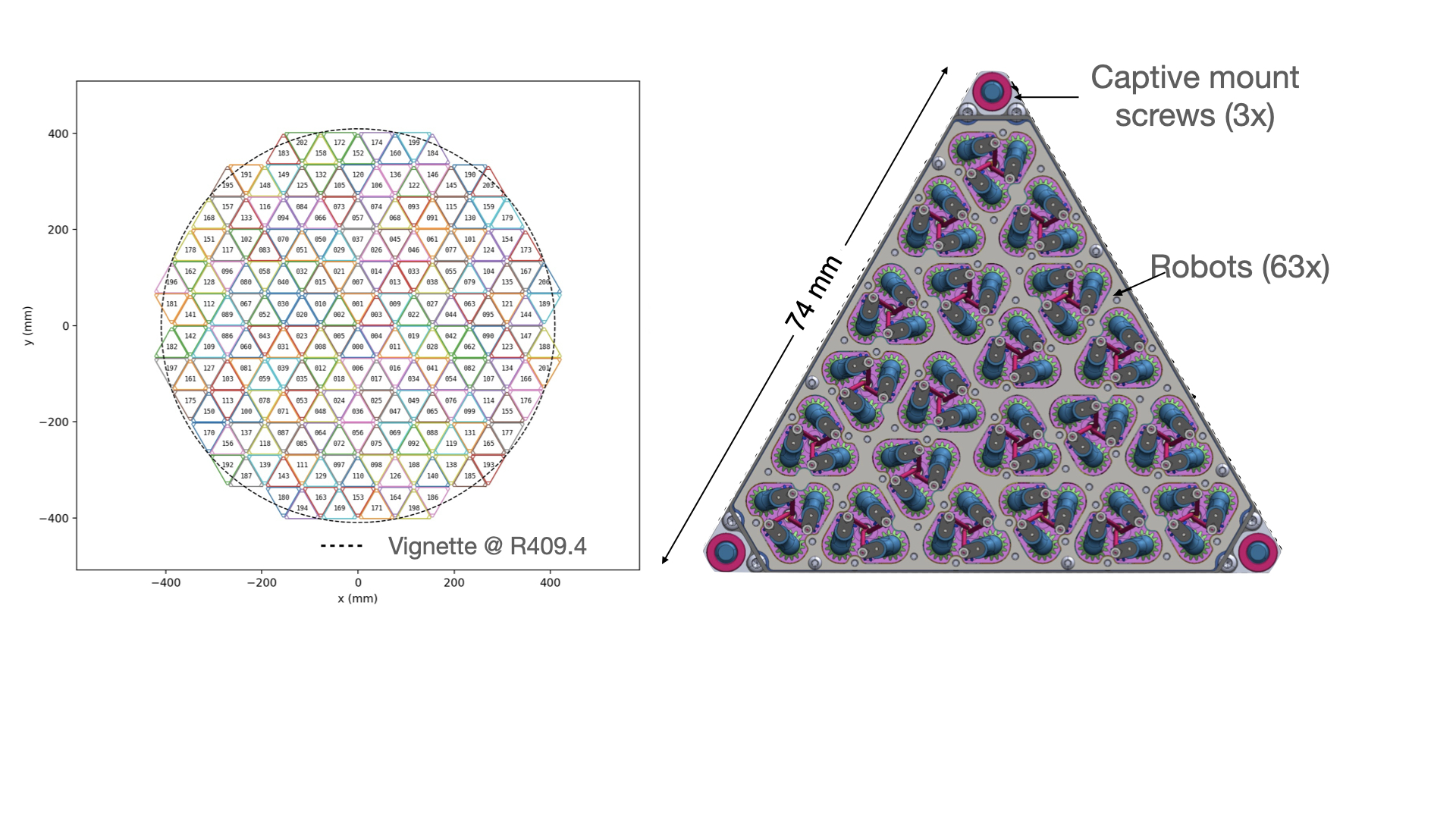}
    \caption{Left: Focal plane layout with 204 rafts. There are 63 fiber positioners per raft resulting in a total of 12,852 fiber positioners on the focal plane. Right: Front view of the raft module, showing how the 63 fiber robots are arranged. In this view we show the Trillium robot units, which position 3 fibers each. The raft design has 3-fold symmetry, and separate electronics boards for groups of 21 robots. }
    \label{fig:focalplane_layout}
\end{figure}

\begin{figure}[th]
    \centering
    \includegraphics[width=5.5in,trim={6cm 10cm 6cm 10cm},clip]{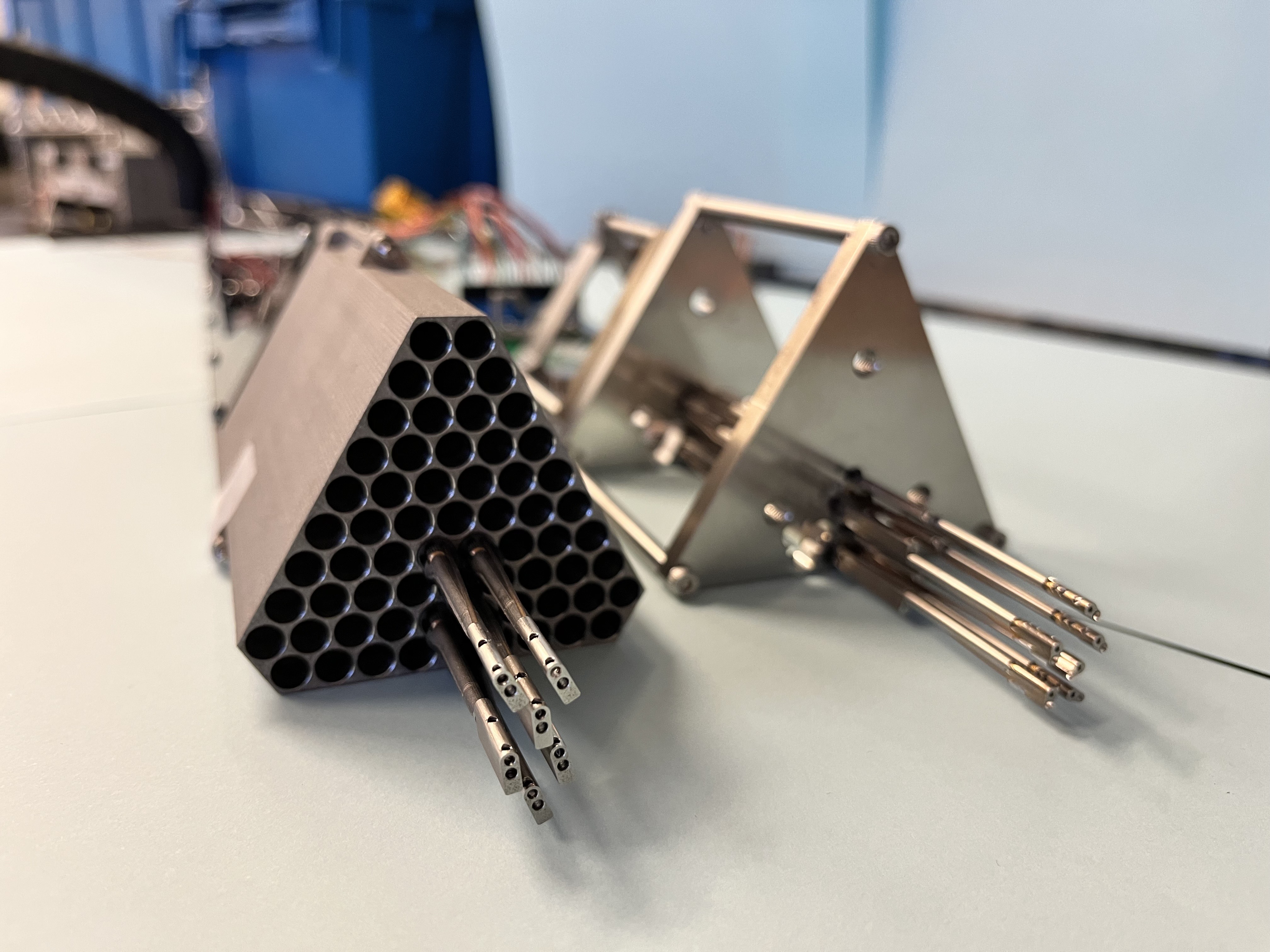}
    \caption{2 protoypes for the raft concept have been manufactured to date. Shell and covers protect the internal components. Clear access is provided for electronics and fiber servicing throughout 2/3 of the length. The 63 fiber robots are divided mechanically and electrically into 3 independent groups of 21, providing further internal symmetry and modularization. The 3 driver electronics boards are particularly easy to replace; each is attached by a single high density connector to its breakout board, to which are soldered the 126 discrete motor wires for those 21 robots.}
    \label{fig:raft_persp}
\end{figure}

The focal plane is composed of a single mechanical structure that accepts the fiber positioners, the guide/focus cameras, fiducial reference sources, and sky monitor fibers. The fiber positioners are modularized in 204 ``rafts'' (Left panel of Figure \ref{fig:focalplane_layout}). Each raft has the shape of an equilateral prism measuring 74 mm on a side and 604 mm long (Right panel of Figure \ref{fig:focalplane_layout}). The raft is a self-contained unit with 63 positioners and control electronics \citep{silber_2022_6354853}. One fiber cable and three Power-over-Ethernet (PoE) cables attach at the rear. The present design weighs only 1.2 kg per raft, making each raft easily removable and serviceable (Figure \ref{fig:raft_persp}).

Fiber positioning robots are mounted at a 6.2 mm center-to-center pitch. This is a significant reduction from the DESI state-of-the-art 10.4 mm minimum pitch. The raft module can accommodate any of several fiber positioning technologies, so long as they can be mounted at 6.2 mm pitch. Our reference design for the raft incorporates 21 ``Trillium'' robot units \citep{silber202225000opticalfiberpositioning}. Each Trillium has 3 fiber actuators, each driven by 2 brushless DC gear motors of diameter 4 mm. These are identical to the motors that drive the DESI robots, and indeed our early prototypes are simply driven by DESI electronics. The change from the DESI design that allowed us to reduce the pitch was the introduction of an additional set of transfer gears which couple the eccentric axis to the central rotation axis. This allows us to nest the motors axially, such that one motor is mounted partially behind the other.

\subsection{Spectrographs}
\label{sec:spectrographs}

Spec-S5 leverages the technology that has enabled the high-performance DESI spectrographs. The total end-to-end throughput of DESI ranges from 30\% -- 50\% for light entering the fibers, which is unprecedented for fiber-fed spectrographs, and a testament to many development efforts and careful design decisions, including the focal ratio and spectrograph optical design. 

Each spectrograph has two dichroics that split the light into three wavelength channels that together record light from 360 -- 980\,nm. The optical design of each channel is optimized for its wavelength range, including the resolution and detector system. The resolution varies from $R = \lambda / \Delta\lambda \approx 2000$ at the shortest wavelengths to $R \approx 5500$ at the longest wavelengths. 

Each site will have 23 spectrographs.  Fibers from eight or nine positioner rafts will be connected to each spectrograph (up to 567 fibers per spectrograph). This is a modest increase over the 500 fibers per spectrograph used by DESI. There will be 36 new spectrographs, as the ten DESI spectrographs will be reused at the Mayall site. Upgrade options include the addition of a 4th arm to the spectrographs. This would increase the wavelength range to $1.2$\,$\mu$m and enable spectroscopy of targets in the redshift desert between $1.5 \leq z \leq 2.0$.

\begin{figure}[th]
    \centering
    \includegraphics[width=6.5in]{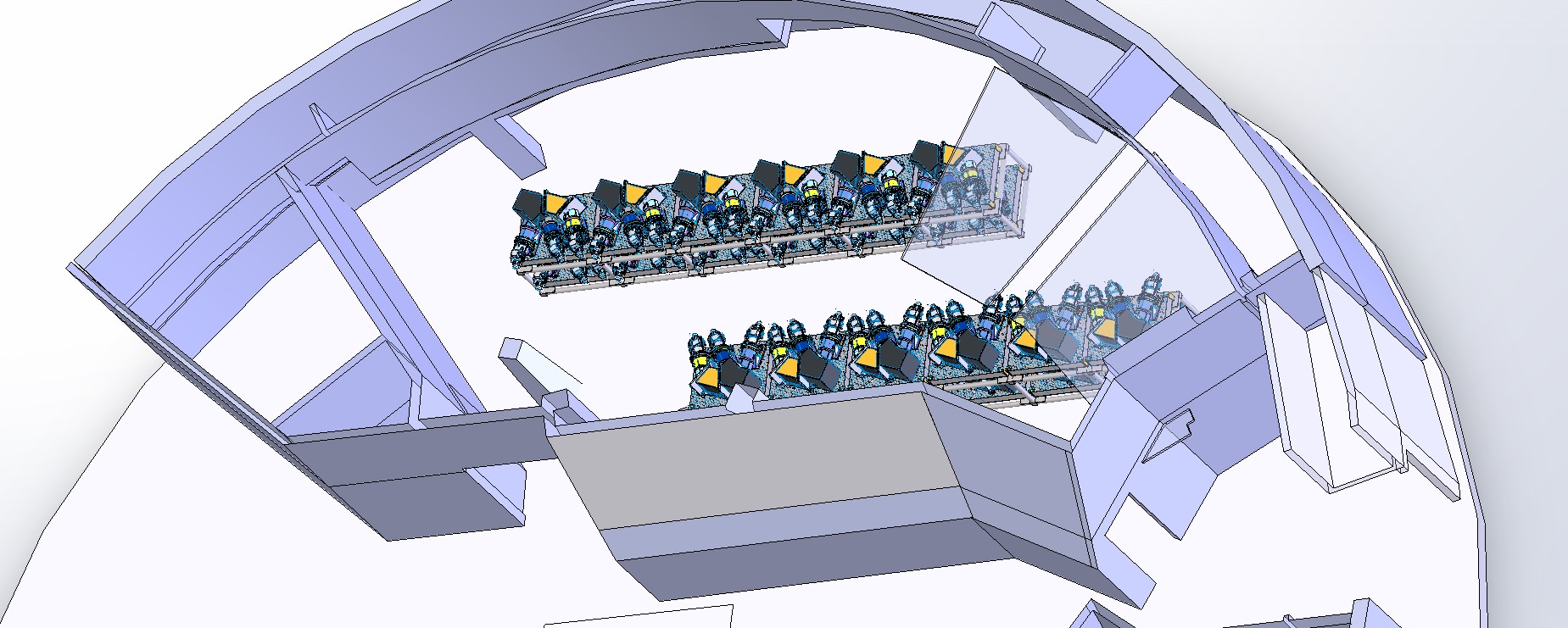}
    \caption{Model that shows the spectrographs in the Large Coud\'e Room at the Mayall telescope mounted on a rack system based on the DESI design. The DESI spectrographs are in a thermal enclosure within the Large Coud\'e Room. The enclosure for Spec-S5 would encompass the entire room. The layout at the Blanco is similar. }
    \label{fig:shack}
\end{figure}

The spectrographs are kept in an environmental enclosure that maintains constant temperature and humidity. This stability plays a key role in the superb calibration and sky subtraction of DESI, and is vital to measure  faint sources. Figure~\ref{fig:shack} shows how the spectrographs would fit within the Large Coud\'e Room at the Mayall on a rack system based on the DESI design. The layout at the Blanco is similar. 

\begin{figure}[th]
    \centering
    \includegraphics[width=\textwidth]{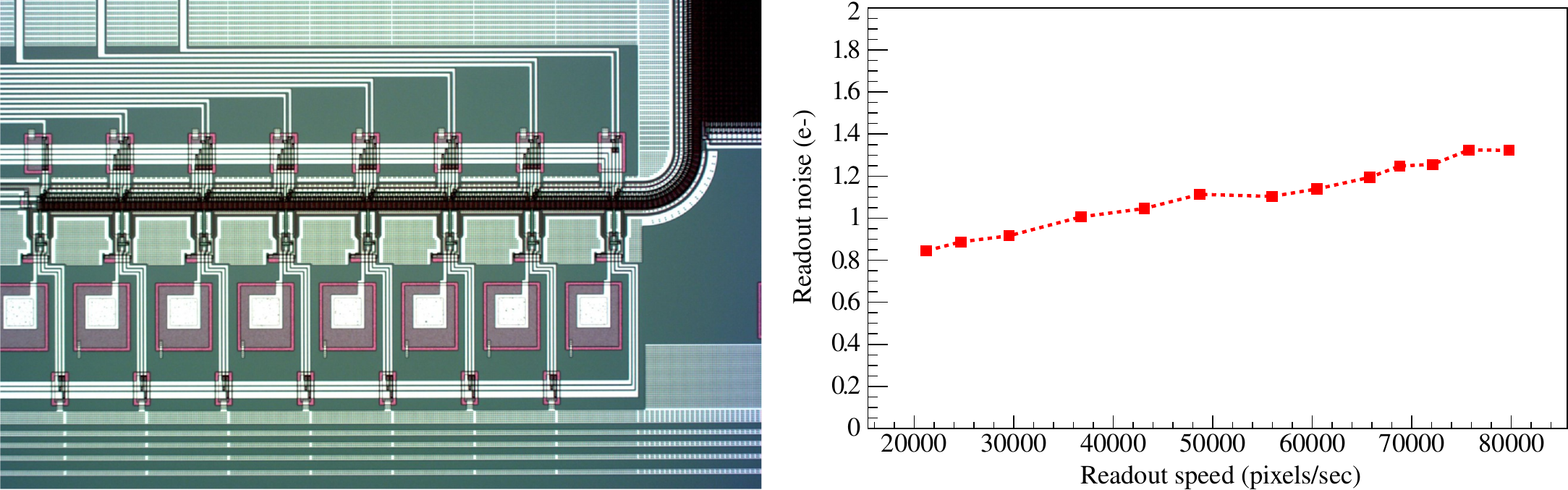}
    \caption{Left: Output region of an 8-channel MAS CCD prototype. Large-format MAS CCDs will enable sub-electron readout noise at comparable readout speed and quantum efficiency as the existing DESI detectors. Figure from \citet{Holland23}. Right: Measured readout noise as a function of the pixel readout speed of a MAS CCD with sixteen output amplifiers using one measurement of the pixel per amplifier.  Figure from \citet{Lapi24}.}
    \label{fig:mas}
\end{figure}

The detector systems for the Spec-S5 spectrographs will utilize novel multi-amplifier sensing (MAS) CCDs (Figure~\ref{fig:mas}) \citep{Holland23, MAS2024} that leverage the non-destructive floating gate readout amplifier system developed for ultra-low-noise Skipper CCDs \citep{Tiffenberg17}. These detectors have been demonstrated to produce dramatically lower read noise ($< 1\,\mathrm{e^-\,rms\,pix^{-1}}$) \citep{MAS2024} than the current DESI CCDs \citep{Bebek17}. 
MAS CCDs have multiple inline amplifiers on an extended serial register, each of which performs an independent measurement of the charge in a pixel. These measurements can be averaged to  reduced the readout noise contribution without the substantial increase in readout time that occurs in conventional Skipper CCDs. 
These detectors will enable the read noise to be substantially subdominant in Spec-S5 observations even at the bluest wavelengths with the lowest sky background. CCDs with Skipper amplifiers have been tested on-sky \citep{Marrufo24}, while prototype MAS CCDs have been extensively tested in the laboratory \citep{MAS2024,Lapi24,Lin24}. Figure~\ref{fig:mas} shows the standard deviation of the readout noise as a function of the pixel readout speed of a MAS CCD after averaging the samples from its sixteen output amplifiers. The design and fabrication of large-format (4k $\times$ 4k) MAS detectors is in progress. The significantly higher channel count of the MAS CCDs will require multiplexing of existing CCD readout electronics, which are currently being developed. The DESI CCDs will be upgraded to this detector technology as part of DESI-2. 

\newpage
\section{Timeline}
\label{sec:Timeline}

\begin{figure}[th]
    \centering
    \includegraphics[width=7in,trim={.1cm .1cm .1cm .1cm},clip]{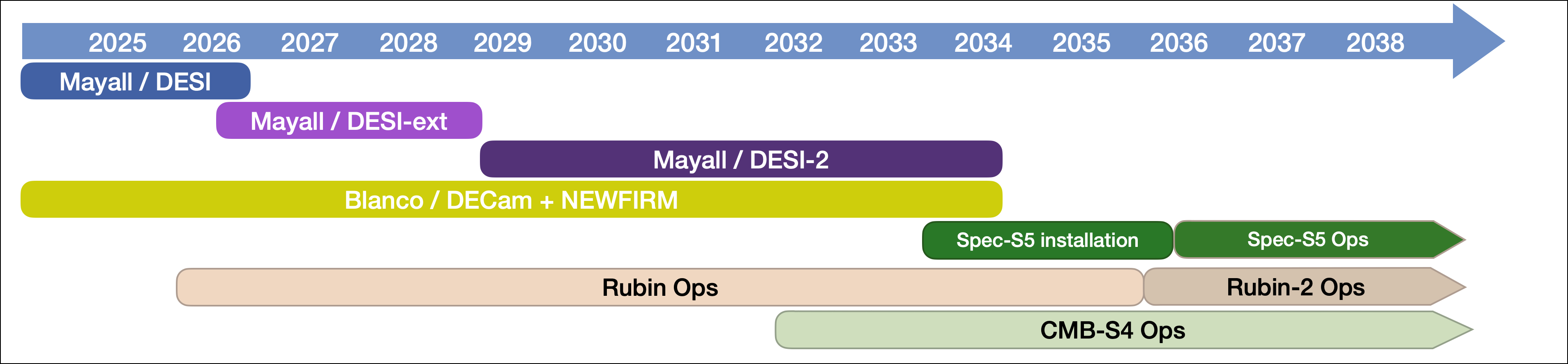}
    \caption{A notional timeline for the design, construction and implementation of Spec-S5. Such an aggressive timeline is only possible using the existing Mayall and Blanco facilities and provides the fastest and most cost-effective path to on-sky operations and executing the planned mission. This timeline also provides the opportunity for Rubin/LSST and an initial phase of Rubin-2 to supply targets for Spec-S5 and for a highly multiplexed spectroscopic capability to support CMB-S4 operations.}
    \label{fig:timeline}
\end{figure}

Figure~\ref{fig:timeline} presents a notional timeline for the Spec-S5 project. This is an aggressive timeline, informed by the experience from the DESI project, the current level of project design and development, an understanding of the state of the facilities, and the requirements of the ongoing projects and commitments on the Mayall and Blanco facilities. 
No significant resources have thus far been allocated to this project, but our Collaboration is actively engaging potential partners and the funding agencies. We welcome broad participation in this project.

\section{Spec-S5 Synergies with Other Facilities}
\label{sec:Synergies}

\subsection{CMB-S4}

The discovery space of high redshift spectroscopic surveys is highly complementary to that of future Cosmic Microwave Background experiments such as CMB-S4 \citep{CMB-S4:2016ple}. For example, the noise in CMB lensing maps will be reduced by more than one order of magnitude over existing measurements.  Unlike cosmic shear, CMB lensing can be measured to very high redshift, providing access to the matter field without the need to model galaxy bias.  Unfortunately CMB lensing alone mostly provides information that is projected along the line-of-sight (and with a broad redshift kernel).  The cross-correlation of CMB lensing with large scale structure (LSS) in several redshift bins (CMB lensing tomography), can break the degeneracy with galaxy bias inherent in LSS alone and provide measurement of the amplitude of perturbations as a function of redshift that cannot be obtained directly from the CMB.  This leads to tighter constraints on neutrino masses and dark energy, while mitigating some of the possible systematics. Moreover, comparing the motion of non-relativistic matter through redshift-space distortions to the deflection of CMB photons will put some of the most informative bounds on theories of modified gravity. At the same time, the cross-correlation of LSS with CMB lensing can potentially improve the robustness of constraints relying on the ultra-large scales, such as measurements of local primordial non-Gaussianity.

In addition to CMB lensing, other secondary anisotropies\footnote{That is, fluctuations caused by the interaction of CMB photons with matter along the line of sight.} of the CMB correlate CMB and LSS maps, revealing the properties of the gas in the cosmic web (through the thermal and kinematic Sunyaev-Zel'dovich effects), and providing a tool to measure large-scale velocities. These velocity measurements can provide access to the very largest scales, often the most affected by primordial physics.

\subsection{Rubin}


The Rubin Observatory LSST is a broad-band imaging survey that will revolutionize our understanding of time-variable and transient astrophysical phenomena, map the matter content of the universe through weak lensing and angular galaxy clustering measurements, measure the (transverse) proper motions of stars, identify new dwarf galaxies and stellar streams around the Milky Way, and use observations of supernovae, strong lenses and galaxy clusters to constrain cosmological parameters \citep{Rubin_LSST_Science_Book_2009}.

Spectroscopy will greatly enhance Rubin's scientific reach \citep{Rubin_LSST_Maximizing_Science_2016}. The Rubin cosmological measurements will rely on accurate photometric redshifts, which in turn require spectroscopic redshift measurements for calibration. In particular, the weak lensing and galaxy clustering studies performed by Rubin will require highly accurate and precise information on the ensemble redshift distributions for the faint galaxy samples used for those measurements \citep{DESC_SRD}. Furthermore, accurate redshift information is necessary for the cosmological interpretation of the galaxy cluster abundances, Type Ia supernovae and strong lenses measured with Rubin. 

Spectroscopic observations are also key to identifying the true nature of transients and variables; measuring the radial velocities, physical properties (i.e., effective temperatures and gravities) and elemental abundances of stars which are key to understanding the formation history of the Milky Way and revealing its dark matter substructure; and for probing the diffuse baryons on interstellar and intergalactic scales. Since the nominal Rubin LSST survey will be complete before Spec-S5 comes on line, Spec-S5 will be an ideal instrument for exploring and characterizing key populations of astronomical targets (stars, galaxies, QSOs, lenses, transient hosts, etc.) identified by Rubin and other ground- and space-based surveys.

Furthermore, as described in \S~\ref{sec:Targets}, an intermediate-band imaging survey with Rubin beginning after LSST (i.e., labelled ``Rubin-2 Ops" in Figure \ref{fig:timeline}) can also quickly provide a large sample of high-redshift LAE targets for Spec-S5. Such a survey covering 9000~deg$^2$ would only require 202 clear dark nights (that is, feasible over approximately 2 years of elapsed time; see Table~\ref{tab_imaging}) to complete and further increase the synergy between the imaging and spectroscopic facilities. 

\section{Summary}
\label{sec:Summary}

We have described the concept for a new facility, Spec-S5, designed to execute the Stage 5 Spectroscopic program. Spec-S5 will address key science drivers in both the 2014 and 2023 reports from the Particle Physics Projects Prioritization Panels (P5) in understanding inflation and dark energy \citepalias{p5report2014,p5report2023}. 
Spec-S5 additionally offers complementary studies of dark matter as a secondary science goal, also recommended in the 2023 P5 report. The highly multiplexed spectroscopic capability delivered by Spec-S5 is also a key science need identified by Astro2020 \citep{Astro2020}.

The reference design for Spec-S5 consists of two identical platforms providing access to both the northern and southern hemispheres.
The first of these will be a 6-m mirror instrumented with a 12,852 fiber instrument hosted on the  Mayall Telescope at Kitt Peak National Observatory.
The second of these will be an identical instrument hosted on the Blanco Telescope at Cerro Tololo Interamerican Observatory.
The instrumentation, target selection, and analysis build directly upon the heritage of the Stage-4 Dark Energy Spectroscopic Instrument (DESI).
This dual-platform, Stage-5 spectroscopic instrument will pursue tests of fundamental physics in three distinct observational programs in an initial six-year survey.

Spec-S5 will dramatically increase our knowledge of early universe physics by mapping the imprint of primordial features over an enormous volume of the distant universe.  This first component of the Spec-S5 survey will map 62 million galaxies and quasars in the $z>2$ universe.
The clustering of these galaxies will provide measurements of large-scale structure that offer an order of magnitude more information content than the ongoing state-of-the-art survey with DESI. The proposed high redshift survey will test inflation and early universe physics that are identified by the 2023 P5 report as the primary science drivers for a new, Stage-5 Spectroscopic experiment. By extending to high redshift, Spec-S5 will constrain the expansion and gravitational growth history in the matter-dominated regime.

The second component of Spec-S5's reference survey will measure more than 100 million galaxies at $z<2$ and expand the DESI observational program to as large an area as possible, thus allowing comprehensive tests of dark energy using the well-established baryon acoustic oscillation (BAO) and redshift space distortion (RSD) techniques. By increasing the survey area beyond the planned DESI program, this new sample will offer the best BAO measurements possible to explore the hints of tension with a $\Lambda$CDM model that have recently appeared in dark energy measurements.

Finally, the third component of Spec-S5's reference survey will pursue new measurements of the distribution of dark matter in the Milky Way on both small and large scales, thus utilizing our host galaxy to test fundamental models of dark matter physics.
In this program, Spec-S5 will be used to provide spectroscopic follow-up observations of stellar streams and dwarf galaxies as recommended in the 2023 P5 report.
By measuring the scale and dynamics of dark matter substructures in the Milky Way's halo, Spec-S5 will probe particle physics models of dark matter that predict observable effects on the halo mass function and/or modify the shapes of dark matter halos.

Spec-S5 is revolutionary for both physics and astronomy. Its initial six-year survey program addresses key science questions from the 2014 and 2023 P5 reports and the Astro2020 report. It will provide an unprecedented spectroscopic database for astrophysics and represents a significant advance relative to the scientific reach of Stage 4 facilities. Spec-S5 will be a critically important facility in the post-Rubin era.



\clearpage

\addcontentsline{toc}{section}{References}

\bibliographystyle{mnras}
\bibliography{bibliography}

\begin{thebibliography}{}
\makeatletter
\relax
\def\mn@urlcharsother{\let\do\@makeother \do\$\do\&\do\#\do\^\do\_\do\%\do\~}
\def\mn@doi{\begingroup\mn@urlcharsother \@ifnextchar [ {\mn@doi@} {\mn@doi@[]}}
\def\mn@doi@[#1]#2{\def\@tempa{#1}\ifx\@tempa\@empty \href {http://dx.doi.org/#2} {doi:#2}\else \href {http://dx.doi.org/#2} {#1}\fi \endgroup}
\def\mn@eprint#1#2{\mn@eprint@#1:#2::\@nil}
\def\mn@eprint@arXiv#1{\href {http://arxiv.org/abs/#1} {{\tt arXiv:#1}}}
\def\mn@eprint@dblp#1{\href {http://dblp.uni-trier.de/rec/bibtex/#1.xml} {dblp:#1}}
\def\mn@eprint@#1:#2:#3:#4\@nil{\def\@tempa {#1}\def\@tempb {#2}\def\@tempc {#3}\ifx \@tempc \@empty \let \@tempc \@tempb \let \@tempb \@tempa \fi \ifx \@tempb \@empty \def\@tempb {arXiv}\fi \@ifundefined {mn@eprint@\@tempb}{\@tempb:\@tempc}{\expandafter \expandafter \csname mn@eprint@\@tempb\endcsname \expandafter{\@tempc}}}

\bibitem[\protect\citeauthoryear{Abazajian et~al.,}{Abazajian et~al.}{2016a}]{abazajian2016cmbs4sciencebookedition}
Abazajian K.~N.,  et~al., 2016a, CMB-S4 Science Book, First Edition (\mn@eprint {arXiv} {1610.02743}), \url {https://arxiv.org/abs/1610.02743}

\bibitem[\protect\citeauthoryear{{Abazajian} et~al.,}{{Abazajian} et~al.}{2016b}]{CMB-S4:2016ple}
{Abazajian} K.~N.,  et~al., 2016b, \mn@doi [arXiv e-prints] {10.48550/arXiv.1610.02743}, \href {https://ui.adsabs.harvard.edu/abs/2016arXiv161002743A} {p. arXiv:1610.02743}

\bibitem[\protect\citeauthoryear{Abdalla et~al.,}{Abdalla et~al.}{2022}]{Abdalla_2022}
Abdalla E.,  et~al., 2022, \mn@doi [Journal of High Energy Astrophysics] {10.1016/j.jheap.2022.04.002}, 34, 49–211

\bibitem[\protect\citeauthoryear{{Abdurro'uf} et~al.,}{{Abdurro'uf} et~al.}{2022}]{2022ApJS..259...35A}
{Abdurro'uf} et~al., 2022, \mn@doi [\apjs] {10.3847/1538-4365/ac4414}, \href {https://ui.adsabs.harvard.edu/abs/2022ApJS..259...35A} {259, 35}

\bibitem[\protect\citeauthoryear{{Adame} et~al.,}{{Adame} et~al.}{2024}]{SV}
{Adame} A.~G.,  et~al., 2024, \mn@doi [\aj] {10.3847/1538-3881/ad0b08}, \href {https://ui.adsabs.harvard.edu/abs/2024AJ....167...62A} {167, 62}

\bibitem[\protect\citeauthoryear{{Adame} et~al.,}{{Adame} et~al.}{2025}]{DESI_BAO_Cosmology_2024}
{Adame} A.~G.,  et~al., 2025, \mn@doi [\jcap] {10.1088/1475-7516/2025/02/021}, \href {https://ui.adsabs.harvard.edu/abs/2025JCAP...02..021A} {2025, 021}

\bibitem[\protect\citeauthoryear{Adams et~al.}{Adams et~al.}{2022}]{Adams:2022pbo}
Adams C.~B.,  et~al., 2022, in {Snowmass 2021}.  (\mn@eprint {arXiv} {2203.14923})

\bibitem[\protect\citeauthoryear{{Alam} et~al.,}{{Alam} et~al.}{2021}]{eBOSS_2021}
{Alam} S.,  et~al., 2021, \mn@doi [\prd] {10.1103/PhysRevD.103.083533}, \href {https://ui.adsabs.harvard.edu/abs/2021PhRvD.103h3533A} {103, 083533}

\bibitem[\protect\citeauthoryear{{Albrecht} et~al.,}{{Albrecht} et~al.}{2006}]{DETF_Report_2006}
{Albrecht} A.,  et~al., 2006, \mn@doi [arXiv e-prints] {10.48550/arXiv.astro-ph/0609591}, \href {https://ui.adsabs.harvard.edu/abs/2006astro.ph..9591A} {pp astro--ph/0609591}

\bibitem[\protect\citeauthoryear{Alibay et~al.,}{Alibay et~al.}{2023}]{10115793}
Alibay F.,  et~al., 2023, in 2023 IEEE Aerospace Conference. pp 1--18, \mn@doi{10.1109/AERO55745.2023.10115793}

\bibitem[\protect\citeauthoryear{{Anderson} et~al.,}{{Anderson} et~al.}{2012}]{BOSS_2012}
{Anderson} L.,  et~al., 2012, \mn@doi [\mnras] {10.1111/j.1365-2966.2012.22066.x}, \href {https://ui.adsabs.harvard.edu/abs/2012MNRAS.427.3435A} {427, 3435}

\bibitem[\protect\citeauthoryear{{Bacon} et~al.,}{{Bacon} et~al.}{2024}]{WST_Overview2024}
{Bacon} R.,  et~al., 2024, \mn@doi [arXiv e-prints] {10.48550/arXiv.2405.12518}, \href {https://ui.adsabs.harvard.edu/abs/2024arXiv240512518B} {p. arXiv:2405.12518}

\bibitem[\protect\citeauthoryear{{Bebek} et~al.,}{{Bebek} et~al.}{2017}]{Bebek17}
{Bebek} C.~J.,  et~al., 2017, \mn@doi [Journal of Instrumentation] {10.1088/1748-0221/12/04/C04018}, \href {https://ui.adsabs.harvard.edu/abs/2017JInst..12C4018B} {12, C04018}

\bibitem[\protect\citeauthoryear{Bechtol et~al.,}{Bechtol et~al.}{2023}]{snowmassDM22}
Bechtol K.,  et~al., 2023, Snowmass2021 Cosmic Frontier White Paper: Dark Matter Physics from Halo Measurements (\mn@eprint {arXiv} {2203.07354}), \url {https://arxiv.org/abs/2203.07354}

\bibitem[\protect\citeauthoryear{Beltrán~Jiménez, Bettoni, Figueruelo  \& Teppa~Pannia}{Beltrán~Jiménez et~al.}{2025}]{Beltr_n_Jim_nez_2025}
Beltrán~Jiménez J.,  Bettoni D.,  Figueruelo D.,   Teppa~Pannia F.~A.,  2025, \mn@doi [Physics of the Dark Universe] {10.1016/j.dark.2024.101761}, 47, 101761

\bibitem[\protect\citeauthoryear{{Blanc} et~al.,}{{Blanc} et~al.}{2022}]{Megamapper2022b}
{Blanc} G.~A.,  et~al., 2022, in {Marshall} H.~K.,  {Spyromilio} J.,   {Usuda} T.,  eds,  Society of Photo-Optical Instrumentation Engineers (SPIE) Conference Series Vol. 12182, Ground-based and Airborne Telescopes IX. p. 1218230, \mn@doi{10.1117/12.2625992}

\bibitem[\protect\citeauthoryear{{Bottaro}, {Castorina}, {Costa}, {Redigolo}  \& {Salvioni}}{{Bottaro} et~al.}{2024a}]{2024arXiv240718252B}
{Bottaro} S.,  {Castorina} E.,  {Costa} M.,  {Redigolo} D.,   {Salvioni} E.,  2024a, \mn@doi [arXiv e-prints] {10.48550/arXiv.2407.18252}, \href {https://ui.adsabs.harvard.edu/abs/2024arXiv240718252B} {p. arXiv:2407.18252}

\bibitem[\protect\citeauthoryear{Bottaro, Castorina, Costa, Redigolo  \& Salvioni}{Bottaro et~al.}{2024b}]{Bottaro:2023wkd}
Bottaro S.,  Castorina E.,  Costa M.,  Redigolo D.,   Salvioni E.,  2024b, \mn@doi [Phys. Rev. Lett.] {10.1103/PhysRevLett.132.201002}, 132, 201002

\bibitem[\protect\citeauthoryear{Botti et~al.,}{Botti et~al.}{2024}]{MAS2024}
Botti A.~M.,  et~al., 2024, in Holland A.~D.,  Minoglou K.,  eds,  SPIE Proceedings Vol. 13103, X-Ray, Optical, and Infrared Detectors for Astronomy XI. SPIE, p. 1310311, \mn@doi{10.1117/12.3019411}, \url {https://doi.org/10.1117/12.3019411}

\bibitem[\protect\citeauthoryear{{Bricman} \& {Gomboc}}{{Bricman} \& {Gomboc}}{2020}]{Bricman2020}
{Bricman} K.,  {Gomboc} A.,  2020, \mn@doi [\apj] {10.3847/1538-4357/ab6989}, \href {https://ui.adsabs.harvard.edu/abs/2020ApJ...890...73B} {890, 73}

\bibitem[\protect\citeauthoryear{{Brout} et~al.,}{{Brout} et~al.}{2022}]{brout22}
{Brout} D.,  et~al., 2022, \mn@doi [\apj] {10.3847/1538-4357/ac8e04}, \href {https://ui.adsabs.harvard.edu/abs/2022ApJ...938..110B} {938, 110}

\bibitem[\protect\citeauthoryear{{Bullock} \& {Boylan-Kolchin}}{{Bullock} \& {Boylan-Kolchin}}{2017}]{BullockBoylanKolchin2017}
{Bullock} J.~S.,  {Boylan-Kolchin} M.,  2017, \mn@doi [\araa] {10.1146/annurev-astro-091916-055313}, \href {https://ui.adsabs.harvard.edu/abs/2017ARA&A..55..343B} {55, 343}

\bibitem[\protect\citeauthoryear{{Burleigh} et~al.,}{{Burleigh} et~al.}{2020}]{burleigh20}
{Burleigh} K.~J.,  et~al., 2020, \mn@doi [\aj] {10.3847/1538-3881/ab93b9}, \href {https://ui.adsabs.harvard.edu/abs/2020AJ....160...61B} {160, 61}

\bibitem[\protect\citeauthoryear{{Cabass}, {Ivanov}, {Philcox}, {Simonovi{\'c}}  \& {Zaldarriaga}}{{Cabass} et~al.}{2023}]{Cabass_2023}
{Cabass} G.,  {Ivanov} M.~M.,  {Philcox} O. H.~E.,  {Simonovi{\'c}} M.,   {Zaldarriaga} M.,  2023, \mn@doi [Physics Letters B] {10.1016/j.physletb.2023.137912}, \href {https://ui.adsabs.harvard.edu/abs/2023PhLB..84137912C} {841, 137912}

\bibitem[\protect\citeauthoryear{{Chou} et~al.,}{{Chou} et~al.}{2022}]{snowmassCF}
{Chou} A.~S.,  et~al., 2022, \mn@doi [arXiv e-prints] {10.48550/arXiv.2211.09978}, \href {https://ui.adsabs.harvard.edu/abs/2022arXiv221109978C} {p. arXiv:2211.09978}

\bibitem[\protect\citeauthoryear{{Cooper} et~al.,}{{Cooper} et~al.}{2023}]{cooper23}
{Cooper} A.~P.,  et~al., 2023, \mn@doi [\apj] {10.3847/1538-4357/acb3c0}, \href {https://ui.adsabs.harvard.edu/abs/2023ApJ...947...37C} {947, 37}

\bibitem[\protect\citeauthoryear{{Czerny} et~al.,}{{Czerny} et~al.}{2023}]{Czerny2023}
{Czerny} B.,  et~al., 2023, \mn@doi [\aap] {10.1051/0004-6361/202345844}, \href {https://ui.adsabs.harvard.edu/abs/2023A&A...675A.163C} {675, A163}

\bibitem[\protect\citeauthoryear{{DES Collaboration} et~al.,}{{DES Collaboration} et~al.}{2024}]{DESY5SN}
{DES Collaboration} et~al., 2024, \mn@doi [\apjl] {10.3847/2041-8213/ad6f9f}, \href {https://ui.adsabs.harvard.edu/abs/2024ApJ...973L..14D} {973, L14}

\bibitem[\protect\citeauthoryear{{DESI Collaboration} et~al.,}{{DESI Collaboration} et~al.}{2016}]{DESI2016_Part1}
{DESI Collaboration} et~al., 2016, \mn@doi [arXiv e-prints] {10.48550/arXiv.1611.00036}, \href {https://ui.adsabs.harvard.edu/abs/2016arXiv161100036D} {p. arXiv:1611.00036}

\bibitem[\protect\citeauthoryear{{DESI Collaboration} et~al.,}{{DESI Collaboration} et~al.}{2022}]{desi_inst}
{DESI Collaboration} et~al., 2022, \mn@doi [\aj] {10.3847/1538-3881/ac882b}, \href {https://ui.adsabs.harvard.edu/abs/2022AJ....164..207D} {164, 207}

\bibitem[\protect\citeauthoryear{{DESI Collaboration} et~al.,}{{DESI Collaboration} et~al.}{2024}]{DESI_FullShape_2024}
{DESI Collaboration} et~al., 2024, arXiv e-prints, \href {https://ui.adsabs.harvard.edu/abs/2024arXiv241112022D} {p. arXiv:2411.12022}

\bibitem[\protect\citeauthoryear{{Davis} et~al.,}{{Davis} et~al.}{2024}]{Davis2024}
{Davis} M.~C.,  et~al., 2024, \mn@doi [\apj] {10.3847/1538-4357/ad276e}, \href {https://ui.adsabs.harvard.edu/abs/2024ApJ...965...34D} {965, 34}

\bibitem[\protect\citeauthoryear{{Dawson} et~al.,}{{Dawson} et~al.}{2013}]{BOSS_2013}
{Dawson} K.~S.,  et~al., 2013, \mn@doi [\aj] {10.1088/0004-6256/145/1/10}, \href {https://ui.adsabs.harvard.edu/abs/2013AJ....145...10D} {145, 10}

\bibitem[\protect\citeauthoryear{{Dawson} et~al.,}{{Dawson} et~al.}{2016}]{eBOSS_2016}
{Dawson} K.~S.,  et~al., 2016, \mn@doi [\aj] {10.3847/0004-6256/151/2/44}, \href {https://ui.adsabs.harvard.edu/abs/2016AJ....151...44D} {151, 44}

\bibitem[\protect\citeauthoryear{{Drlica-Wagner} et~al.,}{{Drlica-Wagner} et~al.}{2022}]{snowmassCF3}
{Drlica-Wagner} A.,  et~al., 2022, \mn@doi [arXiv e-prints] {10.48550/arXiv.2209.08215}, \href {https://ui.adsabs.harvard.edu/abs/2022arXiv220908215D} {p. arXiv:2209.08215}

\bibitem[\protect\citeauthoryear{Dvorkin et~al.}{Dvorkin et~al.}{2022}]{Dvorkin:2022jyg}
Dvorkin C.,  et~al., 2022, in {Snowmass 2021}.  (\mn@eprint {arXiv} {2203.07943})

\bibitem[\protect\citeauthoryear{Ebina \& White}{Ebina \& White}{2024}]{Ebina_2024}
Ebina H.,  White M.,  2024, \mn@doi [Journal of Cosmology and Astroparticle Physics] {10.1088/1475-7516/2024/06/052}, 2024, 052

\bibitem[\protect\citeauthoryear{{Euclid Collaboration} et~al.,}{{Euclid Collaboration} et~al.}{2024}]{euclidoverview}
{Euclid Collaboration} et~al., 2024, \mn@doi [arXiv e-prints] {10.48550/arXiv.2405.13491}, \href {https://ui.adsabs.harvard.edu/abs/2024arXiv240513491E} {p. arXiv:2405.13491}

\bibitem[\protect\citeauthoryear{{Feng}, {Li}, {Zheng}  \& {Tsai}}{{Feng} et~al.}{2020}]{Feng2020}
{Feng} Y.,  {Li} D.,  {Zheng} Z.,   {Tsai} C.-W.,  2020, \mn@doi [\prd] {10.1103/PhysRevD.102.023014}, \href {https://ui.adsabs.harvard.edu/abs/2020PhRvD.102b3014F} {102, 023014}

\bibitem[\protect\citeauthoryear{{Ferraro}, {Sailer}, {Slosar}  \& {White}}{{Ferraro} et~al.}{2022}]{ferraro22}
{Ferraro} S.,  {Sailer} N.,  {Slosar} A.,   {White} M.,  2022, \mn@doi [arXiv e-prints] {10.48550/arXiv.2203.07506}, \href {https://ui.adsabs.harvard.edu/abs/2022arXiv220307506F} {p. arXiv:2203.07506}

\bibitem[\protect\citeauthoryear{Flauger, McAllister, Pajer, Westphal  \& Xu}{Flauger et~al.}{2010}]{flauger10}
Flauger R.,  McAllister L.,  Pajer E.,  Westphal A.,   Xu G.,  2010, \mn@doi [JCAP] {10.1088/1475-7516/2010/06/009}, 06, 009

\bibitem[\protect\citeauthoryear{{Flaugher} et~al.,}{{Flaugher} et~al.}{2015}]{DECam2015}
{Flaugher} B.,  et~al., 2015, \mn@doi [\aj] {10.1088/0004-6256/150/5/150}, \href {https://ui.adsabs.harvard.edu/abs/2015AJ....150..150F} {150, 150}

\bibitem[\protect\citeauthoryear{{Floris}, {Marziani}, {Panda}, {Sniegowska}, {D'Onofrio}, {Deconto-Machado}, {del Olmo}  \& {Czerny}}{{Floris} et~al.}{2024}]{Floris2024}
{Floris} A.,  {Marziani} P.,  {Panda} S.,  {Sniegowska} M.,  {D'Onofrio} M.,  {Deconto-Machado} A.,  {del Olmo} A.,   {Czerny} B.,  2024, \mn@doi [\aap] {10.1051/0004-6361/202450458}, \href {https://ui.adsabs.harvard.edu/abs/2024A&A...689A.321F} {689, A321}

\bibitem[\protect\citeauthoryear{{Gaia Collaboration} et~al.,}{{Gaia Collaboration} et~al.}{2018}]{GaiaDR2}
{Gaia Collaboration} et~al., 2018, \mn@doi [\aap] {10.1051/0004-6361/201833051}, \href {https://ui.adsabs.harvard.edu/abs/2018A&A...616A...1G} {616, A1}

\bibitem[\protect\citeauthoryear{{Gaia Collaboration} et~al.,}{{Gaia Collaboration} et~al.}{2021}]{gaia21}
{Gaia Collaboration} et~al., 2021, \mn@doi [\aap] {10.1051/0004-6361/202039657}, \href {https://ui.adsabs.harvard.edu/abs/2021A&A...649A...1G} {649, A1}

\bibitem[\protect\citeauthoryear{{Gebhardt} et~al.,}{{Gebhardt} et~al.}{2021}]{HETDEX_SurveyDesign}
{Gebhardt} K.,  et~al., 2021, \mn@doi [\apj] {10.3847/1538-4357/ac2e03}, \href {https://ui.adsabs.harvard.edu/abs/2021ApJ...923..217G} {923, 217}

\bibitem[\protect\citeauthoryear{{Graham}, {Connolly}, {Ivezi{\'c}}, {Schmidt}, {Jones}, {Juri{\'c}}, {Daniel}  \& {Yoachim}}{{Graham} et~al.}{2018}]{Graham2018}
{Graham} M.~L.,  {Connolly} A.~J.,  {Ivezi{\'c}} {\v{Z}}.,  {Schmidt} S.~J.,  {Jones} R.~L.,  {Juri{\'c}} M.,  {Daniel} S.~F.,   {Yoachim} P.,  2018, \mn@doi [\aj] {10.3847/1538-3881/aa99d4}, \href {https://ui.adsabs.harvard.edu/abs/2018AJ....155....1G} {155, 1}

\bibitem[\protect\citeauthoryear{{Guiglion} et~al.,}{{Guiglion} et~al.}{2019}]{4MOST_SurveyStrategy2019}
{Guiglion} G.,  et~al., 2019, \mn@doi [The Messenger] {10.18727/0722-6691/5119}, \href {https://ui.adsabs.harvard.edu/abs/2019Msngr.175...17G} {175, 17}

\bibitem[\protect\citeauthoryear{{Guo} et~al.,}{{Guo} et~al.}{2024}]{Guo2024}
{Guo} W.-J.,  et~al., 2024, \mn@doi [\apjs] {10.3847/1538-4365/ad118a}, \href {https://ui.adsabs.harvard.edu/abs/2024ApJS..270...26G} {270, 26}

\bibitem[\protect\citeauthoryear{{Holland}}{{Holland}}{2023}]{Holland23}
{Holland} S.~E.,  2023, \mn@doi [Astronomische Nachrichten] {10.1002/asna.20230072}, \href {https://ui.adsabs.harvard.edu/abs/2023AN....34430072H} {344, e20230072}

\bibitem[\protect\citeauthoryear{{Ivezi{\'c}} et~al.,}{{Ivezi{\'c}} et~al.}{2019}]{2019ApJ...873..111I}
{Ivezi{\'c}} {\v{Z}}.,  et~al., 2019, \mn@doi [\apj] {10.3847/1538-4357/ab042c}, \href {https://ui.adsabs.harvard.edu/abs/2019ApJ...873..111I} {873, 111}

\bibitem[\protect\citeauthoryear{{Jin} et~al.,}{{Jin} et~al.}{2023}]{LAMOST_QSO2023}
{Jin} J.-J.,  et~al., 2023, \mn@doi [\apjs] {10.3847/1538-4365/acaf89}, \href {https://ui.adsabs.harvard.edu/abs/2023ApJS..265...25J} {265, 25}

\bibitem[\protect\citeauthoryear{Kollmeier et~al.,}{Kollmeier et~al.}{2017}]{kollmeier2017}
Kollmeier J.~A.,  et~al., 2017, SDSS-V: Pioneering Panoptic Spectroscopy (\mn@eprint {arXiv} {1711.03234}), \url {https://arxiv.org/abs/1711.03234}

\bibitem[\protect\citeauthoryear{{Kova{\v{c}}evi{\'c}} et~al.,}{{Kova{\v{c}}evi{\'c}} et~al.}{2022}]{Kovacevic2022}
{Kova{\v{c}}evi{\'c}} A.~B.,  et~al., 2022, \mn@doi [\apjs] {10.3847/1538-4365/ac88ce}, \href {https://ui.adsabs.harvard.edu/abs/2022ApJS..262...49K} {262, 49}

\bibitem[\protect\citeauthoryear{LSST Dark Energy Science~Collaboration et~al.,}{LSST Dark Energy Science~Collaboration et~al.}{2021}]{DESC_SRD}
LSST Dark Energy Science~Collaboration R.~M.,  et~al., 2021, The LSST Dark Energy Science Collaboration (DESC) Science Requirements Document (\mn@eprint {arXiv} {1809.01669}), \url {https://arxiv.org/abs/1809.01669}

\bibitem[\protect\citeauthoryear{{LSST Science Collaboration} et~al.,}{{LSST Science Collaboration} et~al.}{2009}]{Rubin_LSST_Science_Book_2009}
{LSST Science Collaboration} et~al., 2009, \mn@doi [arXiv e-prints] {10.48550/arXiv.0912.0201}, \href {https://ui.adsabs.harvard.edu/abs/2009arXiv0912.0201L} {p. arXiv:0912.0201}

\bibitem[\protect\citeauthoryear{Lapi et~al.,}{Lapi et~al.}{2024}]{Lapi24}
Lapi A.~J.,  et~al., 2024, \mn@doi [Journal of Astronomical Telescopes, Instruments, and Systems] {10.1117/1.JATIS.11.1.011203}, 11, 011203

\bibitem[\protect\citeauthoryear{{Laureijs} et~al.,}{{Laureijs} et~al.}{2011}]{laureijs2011eucliddefinitionstudyreport}
{Laureijs} R.,  et~al., 2011, \mn@doi [arXiv e-prints] {10.48550/arXiv.1110.3193}, \href {https://ui.adsabs.harvard.edu/abs/2011arXiv1110.3193L} {p. arXiv:1110.3193}

\bibitem[\protect\citeauthoryear{{Lin} et~al.,}{{Lin} et~al.}{2024}]{Lin24}
{Lin} K.~W.,  et~al., 2024, \mn@doi [\pasp] {10.1088/1538-3873/ad716c}, \href {https://ui.adsabs.harvard.edu/abs/2024PASP..136i5002L} {136, 095002}

\bibitem[\protect\citeauthoryear{{Mainieri} et~al.,}{{Mainieri} et~al.}{2024}]{WST_Science2024}
{Mainieri} V.,  et~al., 2024, \mn@doi [arXiv e-prints] {10.48550/arXiv.2403.05398}, \href {https://ui.adsabs.harvard.edu/abs/2024arXiv240305398M} {p. arXiv:2403.05398}

\bibitem[\protect\citeauthoryear{{Marrufo Villalpando} et~al.,}{{Marrufo Villalpando} et~al.}{2024}]{Marrufo24}
{Marrufo Villalpando} E.,  et~al., 2024, in {Holland} A.~D.,  {Minoglou} K.,  eds,  Society of Photo-Optical Instrumentation Engineers (SPIE) Conference Series Vol. 13103, X-Ray, Optical, and Infrared Detectors for Astronomy XI. p. 131030F (\mn@eprint {arXiv} {2406.10756}), \mn@doi{10.1117/12.3018342}

\bibitem[\protect\citeauthoryear{{McAllister}, {Silverstein}  \& {Westphal}}{{McAllister} et~al.}{2010}]{mcallister10}
{McAllister} L.,  {Silverstein} E.,   {Westphal} A.,  2010, \mn@doi [\prd] {10.1103/PhysRevD.82.046003}, \href {https://ui.adsabs.harvard.edu/abs/2010PhRvD..82d6003M} {82, 046003}

\bibitem[\protect\citeauthoryear{{Najita} et~al.,}{{Najita} et~al.}{2016}]{Rubin_LSST_Maximizing_Science_2016}
{Najita} J.,  et~al., 2016, \mn@doi [arXiv e-prints] {10.48550/arXiv.1610.01661}, \href {https://ui.adsabs.harvard.edu/abs/2016arXiv161001661N} {p. arXiv:1610.01661}

\bibitem[\protect\citeauthoryear{{National Academies of Sciences, Engineering and Medicine}}{{National Academies of Sciences, Engineering and Medicine}}{2021}]{Astro2020}
{National Academies of Sciences, Engineering and Medicine} 2021, {Pathways to Discovery in Astronomy and Astrophysics for the 2020s}.
The National Academies Press, Washington, DC, \mn@doi{10.17226/26141}

\bibitem[\protect\citeauthoryear{{Panda} \& {{\'S}niegowska}}{{Panda} \& {{\'S}niegowska}}{2024}]{Panda2024}
{Panda} S.,  {{\'S}niegowska} M.,  2024, \mn@doi [\apjs] {10.3847/1538-4365/ad344f}, \href {https://ui.adsabs.harvard.edu/abs/2024ApJS..272...13P} {272, 13}

\bibitem[\protect\citeauthoryear{{Particle Physics Project Prioritization Panel}}{{Particle Physics Project Prioritization Panel}}{2014}]{p5report2014}
{Particle Physics Project Prioritization Panel} 2014, Technical report, {2014 P5 Report: Building for Discovery}, \url {https://www.usparticlephysics.org/2014-p5-report/}.
U.S. Particle Physics, \url {https://www.usparticlephysics.org/2014-p5-report/}

\bibitem[\protect\citeauthoryear{{Particle Physics Project Prioritization Panel}}{{Particle Physics Project Prioritization Panel}}{2023}]{p5report2023}
{Particle Physics Project Prioritization Panel} 2023, Technical report, {2023 P5 Report: Exploring the Quantum Universe}, \url {https://www.usparticlephysics.org/2023-p5-report}.
U.S. Particle Physics, \url {https://www.usparticlephysics.org/2023-p5-report}

\bibitem[\protect\citeauthoryear{{Ricci} \& {Trakhtenbrot}}{{Ricci} \& {Trakhtenbrot}}{2023}]{Ricci2023}
{Ricci} C.,  {Trakhtenbrot} B.,  2023, \mn@doi [Nature Astronomy] {10.1038/s41550-023-02108-4}, \href {https://ui.adsabs.harvard.edu/abs/2023NatAs...7.1282R} {7, 1282}

\bibitem[\protect\citeauthoryear{{Richard} et~al.,}{{Richard} et~al.}{2019}]{4MOST_CosmoSurvey}
{Richard} J.,  et~al., 2019, \mn@doi [The Messenger] {10.18727/0722-6691/5127}, \href {https://ui.adsabs.harvard.edu/abs/2019Msngr.175...50R} {175, 50}

\bibitem[\protect\citeauthoryear{Rogers, Hlo\v{z}ek, Lagu\"e, Ivanov, Philcox, Cabass, Akitsu  \& Marsh}{Rogers et~al.}{2023}]{Rogers:2023ezo}
Rogers K.~K.,  Hlo\v{z}ek R.,  Lagu\"e A.,  Ivanov M.~M.,  Philcox O. H.~E.,  Cabass G.,  Akitsu K.,   Marsh D. J.~E.,  2023, \mn@doi [JCAP] {10.1088/1475-7516/2023/06/023}, 06, 023

\bibitem[\protect\citeauthoryear{{Rubin} et~al.,}{{Rubin} et~al.}{2023}]{rubinetal2023}
{Rubin} D.,  et~al., 2023, \mn@doi [arXiv e-prints] {10.48550/arXiv.2311.12098}, \href {https://ui.adsabs.harvard.edu/abs/2023arXiv231112098R} {p. arXiv:2311.12098}

\bibitem[\protect\citeauthoryear{Sailer, Castorina, Ferraro  \& White}{Sailer et~al.}{2021}]{Sailer_2021}
Sailer N.,  Castorina E.,  Ferraro S.,   White M.,  2021, \mn@doi [Journal of Cosmology and Astroparticle Physics] {10.1088/1475-7516/2021/12/049}, 2021, 049

\bibitem[\protect\citeauthoryear{{Schlafly} \& {Finkbeiner}}{{Schlafly} \& {Finkbeiner}}{2011}]{schlafly11}
{Schlafly} E.~F.,  {Finkbeiner} D.~P.,  2011, \mn@doi [\apj] {10.1088/0004-637X/737/2/103}, \href {https://ui.adsabs.harvard.edu/abs/2011ApJ...737..103S} {737, 103}

\bibitem[\protect\citeauthoryear{Schlegel et~al.,}{Schlegel et~al.}{2022b}]{schlegel2022megamapperstage5spectroscopicinstrument}
Schlegel D.~J.,  et~al., 2022b, The MegaMapper: A Stage-5 Spectroscopic Instrument Concept for the Study of Inflation and Dark Energy (\mn@eprint {arXiv} {2209.04322}), \url {https://arxiv.org/abs/2209.04322}

\bibitem[\protect\citeauthoryear{Schlegel et~al.,}{Schlegel et~al.}{2022a}]{schlegel2022spectroscopicroadmapcosmic}
Schlegel D.~J.,  et~al., 2022a, A Spectroscopic Road Map for Cosmic Frontier: DESI, DESI-II, Stage-5 (\mn@eprint {arXiv} {2209.03585}), \url {https://arxiv.org/abs/2209.03585}

\bibitem[\protect\citeauthoryear{{Schlegel} et~al.,}{{Schlegel} et~al.}{2022c}]{2022arXiv220903585S}
{Schlegel} D.~J.,  et~al., 2022c, \mn@doi [arXiv e-prints] {10.48550/arXiv.2209.03585}, \href {https://ui.adsabs.harvard.edu/abs/2022arXiv220903585S} {p. arXiv:2209.03585}

\bibitem[\protect\citeauthoryear{{Schlegel} et~al.,}{{Schlegel} et~al.}{2022d}]{2022arXiv220904322S}
{Schlegel} D.~J.,  et~al., 2022d, \mn@doi [arXiv e-prints] {10.48550/arXiv.2209.04322}, \href {https://ui.adsabs.harvard.edu/abs/2022arXiv220904322S} {p. arXiv:2209.04322}

\bibitem[\protect\citeauthoryear{{Schlegel} et~al.,}{{Schlegel} et~al.}{2022e}]{Megamapper2022a}
{Schlegel} D.~J.,  et~al., 2022e, \mn@doi [arXiv e-prints] {10.48550/arXiv.2209.04322}, \href {https://ui.adsabs.harvard.edu/abs/2022arXiv220904322S} {p. arXiv:2209.04322}

\bibitem[\protect\citeauthoryear{{Shen} et~al.,}{{Shen} et~al.}{2024}]{Shen2024}
{Shen} Y.,  et~al., 2024, \mn@doi [\apjs] {10.3847/1538-4365/ad3936}, \href {https://ui.adsabs.harvard.edu/abs/2024ApJS..272...26S} {272, 26}

\bibitem[\protect\citeauthoryear{Silber}{Silber}{2022}]{silber_2022_6354853}
Silber J.~H.,  2022, Reference Design for MM Raft Fiber Robot Modules, \mn@doi{10.5281/zenodo.6354853}, \url {https://doi.org/10.5281/zenodo.6354853}

\bibitem[\protect\citeauthoryear{{Silber} et~al.,}{{Silber} et~al.}{2022}]{silber202225000opticalfiberpositioning}
{Silber} J.~H.,  et~al., 2022, \mn@doi [arXiv e-prints] {10.48550/arXiv.2212.07908}, \href {https://ui.adsabs.harvard.edu/abs/2022arXiv221207908S} {p. arXiv:2212.07908}

\bibitem[\protect\citeauthoryear{{Slosar}, {Chen}, {Dvorkin}, {Meerburg}, {Wallisch}, {Green}  \& {Silverstein}}{{Slosar} et~al.}{2019}]{Slosar2019}
{Slosar} A.,  {Chen} X.,  {Dvorkin} C.,  {Meerburg} D.,  {Wallisch} B.,  {Green} D.,   {Silverstein} E.,  2019, \mn@doi [\baas] {10.48550/arXiv.1903.09883}, \href {https://ui.adsabs.harvard.edu/abs/2019BAAS...51c..98S} {51, 98}

\bibitem[\protect\citeauthoryear{{Takada} et~al.,}{{Takada} et~al.}{2014}]{PFS2014_Science}
{Takada} M.,  et~al., 2014, \mn@doi [\pasj] {10.1093/pasj/pst019}, \href {https://ui.adsabs.harvard.edu/abs/2014PASJ...66R...1T} {66, R1}

\bibitem[\protect\citeauthoryear{{The MSE Science Team} et~al.,}{{The MSE Science Team} et~al.}{2019}]{MSE_Science}
{The MSE Science Team} et~al., 2019, \mn@doi [arXiv e-prints] {10.48550/arXiv.1904.04907}, \href {https://ui.adsabs.harvard.edu/abs/2019arXiv190404907T} {p. arXiv:1904.04907}

\bibitem[\protect\citeauthoryear{{Tiffenberg}, {Sofo-Haro}, {Drlica-Wagner}, {Essig}, {Guardincerri}, {Holland}, {Volansky}  \& {Yu}}{{Tiffenberg} et~al.}{2017}]{Tiffenberg17}
{Tiffenberg} J.,  {Sofo-Haro} M.,  {Drlica-Wagner} A.,  {Essig} R.,  {Guardincerri} Y.,  {Holland} S.,  {Volansky} T.,   {Yu} T.-T.,  2017, \mn@doi [\prl] {10.1103/PhysRevLett.119.131802}, \href {https://ui.adsabs.harvard.edu/abs/2017PhRvL.119m1802T} {119, 131802}

\bibitem[\protect\citeauthoryear{{Wang} et~al.,}{{Wang} et~al.}{2022}]{RomanHLS}
{Wang} Y.,  et~al., 2022, \mn@doi [\apj] {10.3847/1538-4357/ac4973}, \href {https://ui.adsabs.harvard.edu/abs/2022ApJ...928....1W} {928, 1}

\bibitem[\protect\citeauthoryear{Yan et~al.,}{Yan et~al.}{2022}]{LAMOST_Overview2022}
Yan H.,  et~al., 2022, \mn@doi [The Innovation] {10.1016/j.xinn.2022.100224}, 3, 100224

\bibitem[\protect\citeauthoryear{{Zeltyn} et~al.,}{{Zeltyn} et~al.}{2024}]{Zeltyn2024}
{Zeltyn} G.,  et~al., 2024, \mn@doi [\apj] {10.3847/1538-4357/ad2f30}, \href {https://ui.adsabs.harvard.edu/abs/2024ApJ...966...85Z} {966, 85}

\bibitem[\protect\citeauthoryear{{Zhao} et~al.,}{{Zhao} et~al.}{2024}]{MUST2024}
{Zhao} C.,  et~al., 2024, arXiv e-prints, \href {https://ui.adsabs.harvard.edu/abs/2024arXiv241107970Z} {p. arXiv:2411.07970}

\bibitem[\protect\citeauthoryear{{Zhou} et~al.,}{{Zhou} et~al.}{2024}]{zhou25}
{Zhou} R.,  et~al., 2024, \mn@doi [arXiv e-prints] {10.48550/arXiv.2409.05140}, \href {https://ui.adsabs.harvard.edu/abs/2024arXiv240905140Z} {p. arXiv:2409.05140}

\bibitem[\protect\citeauthoryear{{Zwicky}}{{Zwicky}}{1933}]{Zwicky1933}
{Zwicky} F.,  1933, Helvetica Physica Acta, \href {https://ui.adsabs.harvard.edu/abs/1933AcHPh...6..110Z} {6, 110}

\bibitem[\protect\citeauthoryear{{Zwicky}}{{Zwicky}}{2009}]{Zwicky1933Eng2009}
{Zwicky} F.,  2009, \mn@doi [General Relativity and Gravitation] {10.1007/s10714-008-0707-4}, \href {https://ui.adsabs.harvard.edu/abs/2009GReGr..41..207Z} {41, 207}

\makeatother
\end{thebibliography}

\end{document}